%% file: main.tex
\PassOptionsToPackage{prologue,dvipsnames}{xcolor}
\documentclass[sigconf]{acmart}

\AtBeginDocument{%
  \providecommand\BibTeX{{%
    \normalfont B\kern-0.5em{\scshape i\kern-0.25em b}\kern-0.8em\TeX}}}

\copyrightyear{2024}
\acmYear{2024}
\setcopyright{rightsretained}
\acmConference[CHI '24]{Proceedings of the CHI Conference on Human Factors in Computing Systems}{May 11--16, 2024}{Honolulu, HI, USA}
\acmBooktitle{Proceedings of the CHI Conference on Human Factors in Computing Systems (CHI '24), May 11--16, 2024, Honolulu, HI, USA}
\acmDOI{10.1145/3613904.3642886}
\acmISBN{979-8-4007-0330-0/24/05}
\usepackage{multicol}
\usepackage{multirow}
\usepackage{xspace}
\usepackage[dvipsnames]{xcolor}
\usepackage{makecell}
\usepackage{arydshln}
\usepackage{tabularray}
\usepackage{framed,enumitem} 
\usepackage{tikz}
\usepackage{pifont}

\definecolor{Issue1}{HTML}{B924A5} % magenta
\definecolor{Issue2}{HTML}{BF1656} % pink
\definecolor{Issue3}{HTML}{4860cc} % blue
\definecolor{Issue4}{HTML}{34217A} % purple
\definecolor{Issue5}{HTML}{08703D} % emerald green
\definecolor{Issue6}{HTML}{D76000} % orange

\begin{document}

\title{Demystifying Tacit Knowledge in Graphic Design: Characteristics, Instances, Approaches, and Guidelines}


\author{Kihoon Son}
\email{kihoon.son@kaist.ac.kr}
\affiliation{%
  \institution{School of Computing, KAIST}
  \city{Daejeon}
  \country{Republic of Korea}
}

\author{DaEun Choi}
\email{daeun.choi@kaist.ac.kr}
\affiliation{%
  \institution{School of Computing, KAIST}
  \city{Daejeon}
  \country{Republic of Korea}
}

\author{Tae Soo Kim}
\email{taesoo.kim@kaist.ac.kr}
\affiliation{%
  \institution{School of Computing, KAIST}
  \city{Daejeon}
  \country{Republic of Korea}
}

\author{Juho Kim}
\email{juhokim@kaist.ac.kr}
\affiliation{%
  \institution{School of Computing, KAIST}
  \city{Daejeon}
  \country{Republic of Korea}
}
% Author name shortcut
\renewcommand{\shortauthors}{Kihoon Son, DaEun Choi, Tae Soo Kim, and Juho Kim}

\begin{abstract}
\input{sections/0_abstract.tex}
\end{abstract}






\definecolor{color1}{HTML}{eafdd3}
\definecolor{color2}{HTML}{d3eafd}

\definecolor{colorElement}{HTML}{9966CC}
\definecolor{colorAction}{HTML}{C1E1C1}
\definecolor{colorPurpose}{HTML}{F2D2BD}

% \definecolor{colorCharacteristic}{HTML}{A1CAF1}
% \definecolor{colorInstance}{HTML}{C1E1C1}
% \definecolor{colorApproach}{HTML}{F2D2BD}

\definecolor{colorCharacteristic}{HTML}{F1948A}
\definecolor{colorInstance}{HTML}{76D7C4}
\definecolor{colorApproach}{HTML}{85C1E9}
\definecolor{colorDG}{HTML}{FFC300}

\newcommand{\errCommonsense[1]}{\sethlcolor{color3}{\textbf{\hl{~#1~}}}}
\newcommand{\nameCommonsense}{\errCommonsense[Commonsense]}

% colorbox
\newcommand{\action}[1]{\colorbox{colorAction}{action}}
\newcommand{\element}[1]{\colorbox{colorElement}{element}}
\newcommand{\elements}[1]{\colorbox{colorElement}{elements}}
\newcommand{\purpose}[1]{\colorbox{colorPurpose}{purpose}}

% color circle
% \newcommand{\actionC}[1]{\ding{108} {}}
\newcommand{\actionC}[1]{\tikz\draw[colorAction,fill=colorAction] (0,0) circle (.5ex); {}}
\newcommand{\elementC}[1]{\tikz\draw[colorElement,fill=colorElement] (0,0) circle (.5ex); {}}
\newcommand{\purposeC}[1]{\tikz\draw[colorPurpose,fill=colorPurpose] (0,0) circle (.5ex); {}}

\newcommand{\cR}[1]{\tikz[baseline=-0.7ex]\node[inner sep=3pt, shape=rectangle, draw, minimum width=3pt, colorCharacteristic, fill=colorCharacteristic]{};}
\newcommand{\iR}[1]{\tikz[baseline=-0.7ex]\node[inner sep=3pt, shape=rectangle, draw, minimum width=3pt, colorInstance, fill=colorInstance]{};}
\newcommand{\aR}[1]{\tikz[baseline=-0.7ex]\node[inner sep=3pt, shape=rectangle, draw, minimum width=3pt, colorApproach, fill=colorApproach]{};}
\newcommand{\dgR}[1]{\tikz[baseline=-0.7ex]\node[inner sep=3pt, shape=rectangle, draw, minimum width=3pt, colorDG, fill=colorDG]{};}

\maketitle
\input{sections/1_introduction}

\input{sections/2_related-work}

\input{sections/3_method-overview}

\input{sections/4_characteristics}
\input{sections/5_interview-study}
\input{sections/6_systematic-literature-review}

\input{sections/7_design-guidelines}
\input{sections/8_discussion}
\input{sections/9_conclusion}

\begin{acks}
This work was supported by Institute of Information \& Communications Technology Planning \& Evaluation (IITP) grant funded by the Korea government (MSIT) (No.2021-0-01347,Video Interaction Technologies Using Object-Oriented Video Modeling).
This work was also supported by Institute of Information \& communications Technology Planning \& Evaluation (IITP) grant funded by the Korea government(MSIT) (No.2019-0-00075, Artificial Intelligence Graduate School Program(KAIST))
We thank all of our participants for engaging positively in our various studies.
We also thank all of the members of KIXLAB for their helpful discussions and constructive feedback.
\end{acks}
\bibliographystyle{ACM-Reference-Format}
\bibliography{references}

\end{document}

%% file: sections/0_abstract.tex
Despite the growing demand for professional graphic design knowledge, the tacit nature of design inhibits knowledge sharing. However, there is a limited understanding on the characteristics and instances of tacit knowledge in graphic design. In this work, we build a comprehensive set of tacit knowledge characteristics through a literature review. Through interviews with 10 professional graphic designers, we collected 123 tacit knowledge instances and labeled their characteristics. By qualitatively coding the instances, we identified the prominent elements, actions, and purposes of tacit knowledge. To identify which instances have been addressed the least, we conducted a systematic literature review of prior system support to graphic design. By understanding the reasons for the lack of support on these instances based on their characteristics, we propose design guidelines for capturing and applying tacit knowledge in design tools. This work takes a step towards understanding tacit knowledge, and how this knowledge can be communicated.

%% file: sections/1_introduction.tex
\vspace{15pt}
\section{Introduction}
There is a substantial demand for professional graphic design knowledge, which is essential for graphic designers to create effective graphic design. Graphic designers actively share their design knowledge with others through how-to videos~\cite{howtovideouiux}, design livestreams~\cite{fraser2019sharing, tharatipyakul2020designing, lu2021streamsketch}, or design discussion forums ~\cite{marlow2014rookie}. Also, as most graphic design work is performed through digital design tools nowadays (e.g., Adobe Photoshop\footnote{https://www.adobe.com/products/photoshop.html}, Adobe Illustrator\footnote{https://www.adobe.com/products/illustrator.html}, Figma\footnote{https://www.figma.com/}, etc.), it is possible to track the designer's actions performed within a design tool and the corresponding output. This implies that professional designers can share their work in detail with novices to enable them to learn from their detailed processes.

However, since design knowledge is inherently \textit{tacit}, it is challenging to capture and share with others a designer's knowledge fully exhibited within a design tool ~\cite{bernal2015role}. Knowledge exists on a continuous spectrum, ranging from tacit to explicit, with tacit knowledge and explicit knowledge possessing precisely opposite characteristics ~\cite{leonard1998role}. Unlike explicit knowledge, tacit knowledge is knowledge that is difficult to communicate verbally or through text and, at the same time, it can only be shared by the person who possesses that knowledge. Examples of explicit knowledge include design theories and widely adopted design rules. Examples of tacit knowledge, on the other hand, encompass skills like ``sensing that the composition of design elements is unnatural given the design context'' or ``conducting sufficient reference research before starting a design.'' Since design knowledge is synthesized through personal traits, practical expertise, and experiences with various project tasks~\cite{bernal2015role}, design knowledge inherently contains diverse tacit knowledge.

To enhance the sharing of design knowledge from individuals who possess tacit knowledge to those without it, it is crucial to comprehensively understand and investigate the presence of tacit knowledge in graphic design. In other words, it is crucial to understand the types of tacit knowledge, the characteristics associated with each type, and how existing design tools support this tacit knowledge. Such an understanding will not only shed light on how tacit knowledge is defined in graphic design but will also aid in developing strategies to capture this knowledge from its owners and disseminate it to those without this expertise.

With this goal, this work aims to 1) preferentially construct a comprehensive set of \textbf{characteristics} (\autoref{tab:terms_definition}-Characteristic) that tacit knowledge can possess based on prior work; 2) gather 123 actual examples of tacit knowledge (referred to as \textbf{instance} in this work; \autoref{tab:terms_definition}-Instance) and their characteristics through interviews and annotation studies with graphic design experts; and 3) extract \textit{element}, \textit{action}, and \textit{purpose} from one tacit knowledge instance (\autoref{tab:terms_definition}-Element, Action, and Purpose) and categorize them to thoroughly understand how tacit knowledge is used in the graphic design field. This analysis shows how the tacit knowledge instance of graphic design is being used through what elements, through what actions of designers, and for what purposes. Subsequently, we systematically reviewed existing graphic design support tools or systems (referred to as \textbf{approach}; \autoref{tab:terms_definition}-Approach) to analyze instances of tacit knowledge that are not adequately supported by these tools and the reasons behind such limitations. Our findings revealed that tacit knowledge in graphic design is most commonly applied to \textit{elements} within designs through \textit{actions} related to cognition and manipulation, with the \textit{purposes} of producing visually appealing designs that consider the audience. Overall, tacit knowledge is highly dependent on personal intuition and expertise, but our findings also showed that it is characterized as shareable with others through non-verbal channels in the design process. Still, existing approaches to support graphic design cannot adequately support tacit knowledge conducted by the designer's actions related to sensing and defining, or the knowledge is used for targeting audiences and visuals.

Following our investigation, we propose that tacit knowledge of graphic design can be effectively passed on to others. To facilitate this, we propose two types of design guidelines: 1) guidelines to capture tacit knowledge from designers who possess it and 2) guidelines to support designers who lack this knowledge in applying tacit knowledge in graphic design tools. The capture guidelines describe what type of designer's actions and information the tool should track within and outside a design tool. These guidelines also suggest how the knowledge can be captured by supporting tools that enable designers to annotate their rationale behind specific design actions and patterns. The application guidelines consist of suggestions related to the design of interactions in tools that support designers in \textit{learning-by-doing} by leveraging captured tacit knowledge. Further, we discuss the designers' personal viewpoints in understanding and interpreting the overall tacit knowledge and design creativity. We also discuss the generalizability of our investigation methods to other domains such as coding, writing, or research.

Our contribution is three-fold:
\begin{itemize}
    \item We analyzed tacit knowledge instances in graphic design according to their element, action, and purpose, deeply understanding how this knowledge exists with its characteristics.
    \item We identified types of tacit knowledge inadequately addressed by existing graphic design support approaches and explained the reasons based on their characteristics.
    \item We demonstrated that tacit knowledge of graphic design could be shared beyond verbal channels and proposed design guidelines for design tools capable of capturing and applying this knowledge.
\end{itemize}

%% file: sections/2_related-work.tex
\section{Related Work}
\subsection{Knowledge Sharing Methods in Design Field}
Design knowledge sharing is a critical process for developing the design profession. In the design field, there are various ways to share design knowledge. Traditionally, designers have shared their design knowledge through design studios~\cite{oxman2004think, wilcox2019design}, critiques~\cite{fischer1993embedding, uluoǧlu2000design}, and feedback ~\cite{xu2014voyant, kotturi2019designers}.
In design communities, designers actively share their knowledge or project experiences through discourse~\cite{marlow2014rookie} that also helps experts build knowledge~\cite{goodwin2015professional}, and the set of design principles shared by experts in this way also helps novice's knowledge-sharing process~\cite{yuan2016almost}.
Recently, various online channels have been formed, allowing designers to share their design knowledge with larger audiences through design live streaming ~\cite{fraser2019sharing, tharatipyakul2020designing, yang2020snapstream, lu2021streamsketch} or self-disclosure~\cite{kou2018you}.

However, it is hard to share design knowledge because it includes much tacit knowledge in nature ~\cite{wong2000tacit}. Visual materials are often utilized to convey clearer design knowledge, but even designers often do not know exactly which part of their knowledge is tacit, why, and how to share the knowledge with others ~\cite{cross1982designerly}. In this respect, before considering how to share or communicate professional design knowledge in the graphic domain, this study aims to identify the instances and characteristics of tacit graphical design knowledge first.

% People everyday interact with graphical designs such as web and mobile interfaces, posters, magazines, or packaging ~\cite{zhao2018characterizes}. Since graphic designs are a major avenue through which we receive and consume information ~\cite{bylinskii2017learning, wang2021research}, the graphic design domain is becoming more important. Accordingly, there is an increasing demand for professional design knowledge to learn how to create graphic design that is visually attractive and effectively convey information to target audiences ~\cite{choi2018fontmatcher, fosco2020predicting, guo2021vinci, zheng2019content}. To respond to this need, various online channels have been formed that allow professional designers to share their design knowledge or know-how with larger audiences---overcoming the limits of traditional means such as design studios \cite{oxman2004think, wilcox2019design} and critiques ~\cite{uluoǧlu2000design}. Designers actively share their knowledge or project experience via design communities ~\cite{marlow2014rookie}, self-disclosure  ~\cite{kou2018you}, or by directly showing their design process through formats like design live streaming ~\cite{fraser2019sharing, tharatipyakul2020designing, yang2020snapstream, lu2021streamsketch}.

\subsection{Tacit Nature of Design Knowledge}

Polanyi ~\cite{polanyi2012personal} proposed the concept of tacit knowledge as one of the natures of human knowledge, which has the opposite characteristics of explicit knowledge. Explicit knowledge can be straightforwardly represented and communicated between people~\cite{collins2010tacit}. On the contrary, tacit knowledge has the following features: gained through experiential learning and reflection~\cite{collins2010tacit, schon1983reflective, sharmin2013reflectionspace}, difficult to sufficiently express through verbal means ~\cite{polanyi2009tacit}, accessible only through the knowing subject ~\cite{lam2000tacit}, deeply embedded in action patterns and external representation~\cite{schon2017reflective, suwa2002external}, and expressed differently depending on the context ~\cite{nonaka1994dynamic}. Tacit knowledge is crucial, as it significantly enhances the excellence of knowledge and its application to tasks.~\cite{seidler2008use, goffee2000should, kirsh1994distinguishing}. Since this knowledge is also an essential part of human knowledge, the HCI community has investigated tacit knowledge to unpack and utilize it ~\cite{prpa2020articulating, kirsh2013embodied}.
% According to Leonard and Sensiper~\cite{leonard1998role}, knowledge exists on a spectrum from tacit, unconscious knowledge to explicit, codified knowledge. The majority of knowledge falls somewhere in this knowledge spectrum. This can be interpreted as having varying degrees of tacitness ~\cite{odell1983discourse, howells1996tacit} depending on how difficult it is to share and communicate. Hence, identifying the characteristics of each tacit knowledge has to be the first step to deeply understanding it.

Design knowledge has many aspects that are tacit in nature ~\cite{bernal2015role}. For example, the design principle is one of explicit knowledge, but knowing the right time to apply the theory is an example of tacit knowledge. In a design process, designers usually make design decisions at the cognitive level ~\cite{lin2020designers, faste2017intuition}. Due to decisions being derived from intuition ~\cite{faste2017intuition} or experience ~\cite{nakakoji1995intertwining}, there are frequent situations in which designers cannot clearly explain what they did ~\cite{cross1982designerly}. This aspect of design knowledge makes sharing or communicating knowledge more challenging in the learning context.

However, although various tacit knowledge was investigated by prior work, there is a limited understanding of what instances of tacit knowledge exist or what characteristics cause the communication process to be difficult. Also, there is no comprehensive investigation of how the prior work on graphic design support approaches covered what instances of tacit knowledge. Therefore, this work focuses on specifying what characteristics and instances of tacit knowledge exist in graphic design.

%% file: sections/3_method-overview.tex
\begin{table*}[!t]
\caption{The definition and examples of each term used in this paper.}
\label{tab:terms_definition}
\def\arraystretch{1.4}%
\resizebox{0.9\linewidth}{!}{%
\begin{tabular}{ccm{\columnwidth}m{0.9\columnwidth}} 
    \cmidrule[1.2pt]{1-4}
    \multicolumn{2}{c}{\textbf{Terms}} &
      \textbf{Definition} &
      \textbf{Example} \\
    \cmidrule[1.2pt]{1-4}
    \multicolumn{2}{c}{\textbf{Characteristic}} &
      Features that tacit knowledge may possess &
      The knowledge that is difficult to articulate verbally \\ \hline
    \multicolumn{2}{c}{\textbf{Instance}} &
      Specific examples of tacit knowledge found in graphic design &
      Harmoniously arranging images to match the background \\
    \cmidrule[0.8pt]{1-4}
    \multirow{3}{*}{\begin{tabular}[c]{@{}c@{}}\textbf{Analytics} \\ \textbf{Schemes}\end{tabular}} &
      \textit{Element} &
      Where tacit knowledge is revealed or manifested in graphic design &
      \textcolor{lightgray}{Harmoniously arranging} \textit{images} \textcolor{lightgray}{to match the background} \\ \cline{2-4}
     &
      \textit{Action} &
      Actions performed by the subject to reveal their tacit knowledge &
      \textcolor{lightgray}{Harmoniously} \textit{arranging} \textcolor{lightgray}{images to match the background} \\ \cline{2-4}
     &
      \textit{Purpose} &
      Purposes for the subject of using their tacit knowledge &
      \textit{Harmoniously} \textcolor{lightgray}{arranging images} \textit{to match the background} \\
    \cmidrule[0.8pt]{1-4}
    \multicolumn{2}{c}{\textbf{Approach}} &
      Methods from prior work supporting the use of tacit knowledge & Vinci: an intelligent graphic design system for generating advertising posters~\cite{guo2021vinci}
       \\ 
\cmidrule[1.2pt]{1-4}
\end{tabular}%
}
\end{table*}

\section{Method Overview and Research Questions}

\subsection{Goals and Research Questions}
Our study aims to investigate tacit knowledge in graphic design deeply by identifying the characteristics and instances of tacit knowledge, as well as by reviewing prior approaches for supporting the usage of design knowledge. The ultimate goal is to propose design guidelines for systems that support the capture and application of tacit knowledge. Based on these research goals, we defined three research questions:
\begin{itemize}
    \item RQ1. What types of tacit knowledge \textbf{instances} exist in graphic design, and what are the differences between these types regarding their \textbf{characteristics}?
    \item RQ2: How do prior \textbf{approaches} in graphic design support the usage of tacit knowledge \textbf{instances}, and what parts lack support?
    \item RQ3. How should tools that support the sharing, communicating, and using tacit knowledge be designed?
\end{itemize}

\subsection{Definition of Terms}
This research employs various terms to analyze tacit knowledge. For clarity, we first provide definitions and examples of these terms in this section. First, the term \textbf{Characteristic} refers to features that tacit knowledge may possess (e.g., knowledge that is \textit{``difficult to articulate verbally''}). Next, the term \textbf{Instance} used in this work refers to specific examples of tacit knowledge found in graphic design, like \textit{``harmoniously arranging images to match the background''}. \textit{Elements}, \textit{actions}, and \textit{purposes} serve as schemes to analyse \textbf{instances} in detail; in the previous example, the \textit{element} is \textit{``image''}, \textit{action} is \textit{``arrange''}, and the \textit{purpose} is \textit{``harmony''}. Utilizing these schemes, we analyzed instances collected from graphic design experts. Finally, the term \textbf{``Approach''} signifies methods from prior work that support the use of tacit knowledge. As these terms are frequently used, we have compiled definitions and examples for each term in \autoref{tab:terms_definition}.

\subsection{Research Procedure Overview}
\begin{figure*}[!ht]
  \centering
  \includegraphics[width=0.8\linewidth]{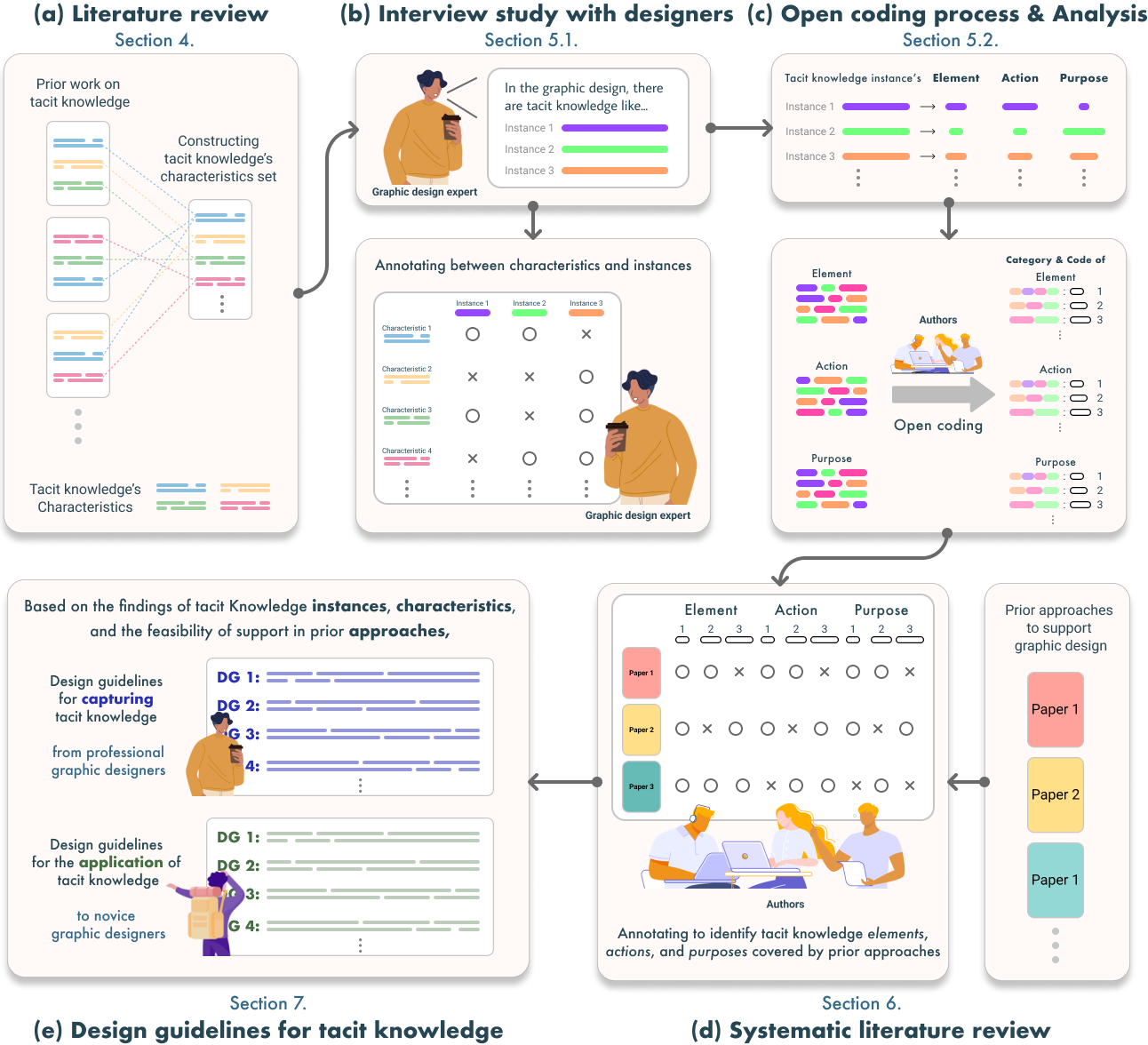}
  \caption{Research overview. The diagram illustrates the overall research process of this work. Each research step is depicted, with specific sections marked corresponding to the description of each stage.}
  \Description{The diagram shows the overall research procedures with arrows. First, (a) is a literature review process for collecting tacit knowledge characteristics. (b) shows that the authors did an interview study with graphic design experts to collect the tacit knowledge instances. In the study, the designers also conducted an annotation task between the instances they mentioned and the characteristics set. (c) is the open coding process to analyze the instance of the collected tacit knowledge. Then, the authors conducted a systematic literature review to investigate what type of instance is still uncovered at (d). Based on these series of research steps, this paper suggests the design guidelines for capturing and applying tacit knowledge at (e).}
  \label{fig:Procedure}
\end{figure*}

Our research consisted of five main steps, which are illustrated in \autoref{fig:Procedure}.
First, we examined the \textbf{characteristics} that are widely used in prior literature to describe tacit knowledge and created an integrative set of these characteristics (\autoref{fig:Procedure}-a). Subsequently, we conducted an interview study with graphic design experts, where we asked them to recollect \textbf{instances} of tacit knowledge and to annotate the characteristics of these instances (\autoref{fig:Procedure}-b). To analyze the collected data in-depth, we also extracted the \textit{elements}, \textit{actions}, and \textit{purposes} from each instance and conducted qualitative coding  (\autoref{fig:Procedure}-c). As a result, we analyzed the \textbf{characteristics}, \textit{elements}, \textit{actions}, and \textit{purposes} associated with the instances. Next, we conducted a systematic literature review to understand why prior \textbf{approaches} may or may not support certain \textbf{instances} due to their specific \textbf{characteristics} (\autoref{fig:Procedure}-d). Through this series of research processes, we ultimately propose \textbf{design guidelines} for supporting the capture and application of tacit knowledge in the graphic design process (\autoref{fig:Procedure}-e).

%% file: sections/4_characteristics.tex
\section{Compiling Characteristics of Tacit Knowledge Through a Literature Review}

\subsection{Tacit Knowledge Characteristics}
According to Leonard and Sensiper~\cite{leonard1998role}, knowledge exists on a spectrum from tacit and unconscious knowledge to explicit and codified knowledge. The majority of knowledge falls somewhere in this knowledge spectrum. This can be interpreted as having varying degrees of tacitness ~\cite{odell1983discourse, howells1996tacit} depending on how difficult it is to share and communicate. Hence, identifying the characteristics of each tacit knowledge has to be the first step to understanding it deeply.

For example, \textit{``sensing that the composition of design elements is unnatural''} could have more intuition-related characteristics than \textit{``conducting sufficient reference research before starting a design''}. In addition, if the former instance is more challenging to share with others than the latter, we can say that the former instance has a higher tacitness than the latter. Therefore, it is crucial to understand these characteristics and how they manifest in different pieces of tacit knowledge to understand the knowledge. While prior work has described some of these characteristics, to the best of our knowledge, no study has provided a comprehensive list with the described characteristics from prior work. In this background, we first assembled and organized the characteristics defined in prior work through a literature survey.
\label{section_characteristic}

\begin{table*}[!ht]
\caption{A comprehensive list of the characteristics of tacit knowledge and their categories.}
\Description{A table shows the result of the literature review on tacit knowledge. The category characteristics and paper of the tacit knowledge are illustrated. There are four categories, and 14 characteristic codes are shown in the table. In category, possession, expression, acquisition, and manifestation exist.}
\label{tab:tacit-characteristics-list}
\begin{center}
\def\arraystretch{1.5}%
\resizebox{0.95\linewidth}{!}{%
\begin{tabular}{cp{1.4\columnwidth}l}
\cmidrule[1.2pt]{1-3}
    \textbf{Category} & \textbf{Characteristic Code} & \textbf{Prior work} \\
\cmidrule[1.2pt]{1-3}
    \multirow{3}{*}{Possession} & Individuals are able to recognize the existence of their tacit knowledge & ~\cite{gertler2003tacit, joia2010relevant, lam2000tacit} \\
    & Individuals possess knowledge that is personal, such as personal skills or routines & ~\cite{ambrosini2001tacit, mcadam2007exploring, erden2008quality, boiral2002tacit, armstrong2008experiential, leonard1998role, seidler2008use, panahi2013towards, desouza2003facilitating, eraut2000non, lam2000tacit} \\
    & Individuals possess knowledge that is practical, such as rules or conventions in practice (e.g., specific to a certain industry or domain) & ~\cite{howells1996tacit, ambrosini2001tacit, mcadam2007exploring, koskinen2002role} \\
\cmidrule[0.5pt]{1-3}
    \multirow{5}{*}{Expression} & Knowledge is expressed in human actions & ~\cite{koskinen2003tacit, seidler2008use, collins2010tacit, odell1983discourse} \\
    & Knowledge is hard to articulate but can be inferred via non-verbal channels\ & ~\cite{mcadam2007exploring, boiral2002tacit, armstrong2008experiential, collins2010tacit, koskinen2002role, goffin2011tacit, k2001tacit} \\
    & Knowledge is not codified and difficult to write down or formalize & ~\cite{howells1996tacit, ambrosini2001tacit, mcadam2007exploring, erden2008quality, boiral2002tacit, gertler2003tacit, seidler2008use, foos2006tacit, desouza2003facilitating, joia2010relevant, holste2010trust, nash2006tacit} \\
    & Knowledge is difficult to explain in terms of the decision rules that were used to reason their application & ~\cite{ambrosini2001tacit, mcadam2007exploring, collins2010tacit, eraut2000non} \\
    & Knowledge is difficult to share or communicate & ~\cite{boiral2002tacit, gertler2003tacit, panahi2013towards, koskinen2002role, johannessen2001mismanagement, haldin2000difficulties, desouza2003facilitating, joia2010relevant, goffin2011tacit, spender1993competitive, k2001tacit} \\
\cmidrule[0.5pt]{1-3}
    \multirow{2}{*}{Acquisition} & Knowledge is implicitly learned by doing and using & ~\cite{howells1996tacit, mcadam2007exploring, koskinen2003tacit, boiral2002tacit, gertler2003tacit, johannessen2001mismanagement, haldin2000difficulties, foos2006tacit, joia2010relevant, mascitelli2000experience, k2001tacit, nash2006tacit} \\
    & The process of acquiring knowledge is affected by personal traits, including mental models, values, beliefs, perceptions, and insights & ~\cite{mcadam2007exploring, koskinen2003tacit, leonard1998role, seidler2008use, koskinen2002role, foos2006tacit, desouza2003facilitating, holste2010trust, k2001tacit} \\
\cmidrule[0.5pt]{1-3}
    \multirow{4}{*}{Manifestation} & Knowledge is specific to the context in which it is used & ~\cite{ambrosini2001tacit, mcadam2007exploring, koskinen2003tacit, boiral2002tacit, johannessen2001mismanagement, holste2010trust, k2001tacit} \\
    & Knowledge is specific to professionals and their experiences & ~\cite{koskinen2003tacit, boiral2002tacit, leonard1998role, holste2010trust, eraut2000non, mascitelli2000experience} \\
    & Knowledge is related to an individual's personal intuition, gut feelings, or experience & ~\cite{erden2008quality, armstrong2008experiential, panahi2013towards, koskinen2002role, haldin2000difficulties, desouza2003facilitating, holste2010trust, spender1993competitive, k2001tacit} \\
    & Knowledge is conventional knowledge that is affected by societal values, tradition, or culture & ~\cite{collins2010tacit} \\
\cmidrule[0.5pt]{1-3}
\end{tabular}
}
\end{center}
\end{table*}

\subsection{Search Keywords and Exclusion Criteria}
To define a comprehensive set of tacit knowledge characteristics within the vast literature, we narrowed our survey by setting limited search keywords. Specifically, we used the search keywords {\itshape ``Tacit knowledge''} and {\itshape ``Definition of tacit knowledge''} to identify literature in Google Scholar\footnote{https://scholar.google.com/} and the ACM Digital Library\footnote{https://dl.acm.org/}. Then, to construct a set based on the most widely used characteristics of tacit knowledge, we only included literature with more than 300 citations with no restrictions on the publication year. Second, we filtered out papers that had no specific sections or parts describing the characteristics of tacit knowledge. Based on this procedure, we first collected 398 papers through the search keywords, and filtered out 321 and 49 papers with the first and second constraints, respectively. As a result, we landed on a set of 28 research papers. We found 13 papers with the first keyword and 15 papers with the second keyword.

\subsection{Coding Process of Tacit Knowledge Characteristics}
From the sections or fragments in the 28 papers that described the characteristics of tacit knowledge, we collected 102 data points related to these characteristics. Specifically, we collected data from a paper in the following situations: 1) when a particular section name includes ``characteristics'', ``features'', or ``definitions'' of tacit knowledge, and 2) when those are explained like ``Tacit knowledge characteristics are A or B''. Two authors conducted an iterative coding process on the data points to merge overlapping characteristics into one code and build a category for similar codes. The authors classified each data point by merging and dividing it into groups. Next, three authors, including another author, met and reviewed the groups and finalized the categorization of code groups. All steps were repeated until the participating authors reached a consensus.

Through this process, we created a list of 14 characteristics with 4 categories in tacit knowledge (\autoref{tab:tacit-characteristics-list}). The categories include ``Possession'',  ``Expression'', ``Acquisition'', and ``Manifestation''. ``Possession'' is related to how to possess tacit knowledge; ``Expression'' is about how to express tacit knowledge; ``Acquisition'' is related to acquiring tacit knowledge one does not have; and ``Manifestation'' is about in what context and situations tacit knowledge is used.

%% file: sections/5_interview-study.tex
\section{Interview Study with Professional Graphic Designers}
To identify concrete instances of tacit knowledge in the graphic design domain and understand what characteristics these instances possess, we conducted an interview study with 10 graphic design experts (Section~\ref{section_interviewStudy}). Through the study, we collected 123 instances of tacit knowledge. Then, we analyzed how tacit knowledge is manifested in graphic design by performing a qualitative coding of the collected instances according to the element where the tacit knowledge is revealed, the action taken by the knowledge subject to use the tacit knowledge, and the purpose of using that knowledge. 
This study and analysis aim to address the first RQ of our work through a set of more detailed RQs:
\begin{itemize}
    \item RQ1. What types of tacit knowledge \textbf{instances} exist in graphic design, and how do they differ regarding tacit knowledge characteristics?
    \begin{itemize}
        \item RQ1-1. Through what \textit{elements}, \textit{actions}, and \textit{purposes} is tacit knowledge in graphic design used?
        \item RQ1-2. In terms of the \textit{element}, what characteristics do instances have where tacit knowledge is revealed?
        \item RQ1-3. In terms of the \textit{action} performed to use the tacit knowledge, what characteristics do instances have?
        \item RQ1-4. What characteristics do instances have regarding the \textit{purpose} of using tacit knowledge?
    \end{itemize}
\end{itemize}
\label{section_rq1}

\subsection{Interview Study}
\label{section_interviewStudy}

\subsubsection{Interviewees and Recruitment}
Since one of the characteristics of tacit knowledge is ``professional'' ~\cite{koskinen2003tacit, boiral2002tacit} and ``experienced'' ~\cite{howells1996tacit, mcadam2007exploring, koskinen2003tacit} knowledge, we recruited graphic design experts with at least five years of experience. A total of 10 participants (\autoref{tab:interviewees}) were recruited through the online communities and social media (Kakao\footnote{https://www.kakaocorp.com} open group chat room or Facebook\footnote{https://www.facebook.com/} group for graphic designers). Specifically, we recruited graphic design experts in BX~\footnote{Brand experience design.}, UI/UX, product, and editorial~\footnote{Closely related to physical printing design like posters, packaging, or magazines.} design. The participants consisted of 3 male and 7 female designers, and their average length of experience was 8.1 years (Min=5, Max=13, and SD=2.96).

\begin{table*}[!ht]
\caption{Information regarding the domain, experience, and type of design work performed by the 10 designers interviewed.}
\Description{Table 2 shows the participant's list and their information. Participant’s ID, gender, age, domain, experience, and their current work and examples of work are illustrated in the table.}
\label{tab:interviewees}
\begin{center}
\resizebox{\linewidth}{!}{%
\def\arraystretch{1.2}%
\begin{tabular}{llllll}
\cmidrule[1.2pt]{1-6}
\textbf{ID} & \textbf{Gender} & \textbf{Age} & \textbf{Domain} & \textbf{Experience} & \textbf{Current work (Examples of work)}\\
\cmidrule[1.2pt]{1-6}
    P1 & Male & 31 & Product & 7 years & B2B service design project manager (from planning to actual service design)\\
    
    P2 & Female & 30 & Product & 5 years & IT product services project manager (planning, designing, prototype testing, and validation of service)\\
    
    P3 & Male & 39 & BX & 13 years & Branding design work related to broadcast media-focused (TV, news, and internet)\\
    
    P4 & Female & 34 & BX & 8 years & Branding design focusing on e-commerce (from packaging to design promotion)\\
    
    P5 & Female & 35 & UI/UX & 10 years & E-commerce service UI/UX design (data analysis-based design or seller side UI/UX design)\\
    
    P6 & Male & 34 & UI/UX & 7 years & In-house UI/UX design (landing page, detailed description page, or company web page design)\\
    
    P7 & Male & 31 & UI/UX & 5 years & UI/UX design in coding education start-up (UI/UX design of the actual page design through communication with developers)\\
    
    P8 & Female & 33 & Editorial& 7 years & Editorial design of various channels (advertisement, signboard, poster, pamphlet, card news, logo, etc.)\\
    
    P9 & Female & 32 & Editorial & 6 years & Kids related editorial design (product packaging, poster, advertisement, including character design)\\
    
    P10 & Female & 36 & Editorial & 13 years & Editorial design of various channels (museum or cosmetics catalog, mobility poster, proposal design for large companies)\\
\cmidrule[1.2pt]{1-6}
\end{tabular}
% }
}
\end{center}
% \vspace{-5mm}
\end{table*}

\subsubsection{Protocol}
Two authors led the study (90 minutes). First, the study started by explaining the definition of tacit knowledge (5 minutes). Next, based on the pre-defined questions in a semi-structured interview format, we asked participants to think of and explain instances of tacit knowledge in the domain they work in (35 minutes). In their work process, we asked whether tacit knowledge exists in their domain, how professional and novice designers differ in using tacit knowledge, and what the specific experience related to tacit knowledge would be. What is needed to capture and apply tacit knowledge in the design process was also asked. During the interview, another author listed the explained instances of tacit knowledge.

After the interview, participants were asked to look through the instances they mentioned and merge similar ones or divide some mixed instances. For example, P10 explained \textit{``A sense of judging strange balance, symmetry, ratio, or color of elements''} as tacit knowledge in graphic design. However, P10 again divided it into \textit{``A sense of judging strange balance, symmetry, ratio of elements''} and \textit{``Defining a good color scheme for a design''} because P10 thought the color was a distinct element compared to others. After that, we asked them to check the proper characteristics for each instance by looking at the characteristics set (\autoref{tab:tacit-characteristics-list}) we built (50 minutes). Participants were compensated with 105,000 KRW (Approx. 80 USD). The studies were conducted remotely through Zoom\footnote{https://zoom.us/}, and the sessions were recorded.

\subsection{Analysis Method and Goal on Tacit Knowledge Instances}
To investigate the collected instances thoroughly, we conducted open coding on each instance and analyzed the characteristics annotated by experts. This section describes the analysis scheme and process of open coding, as well as the annotation result analysis method.

\subsubsection{Analysis Scheme: Element, Action, and Purpose of Tacit Knowledge}
We defined three perspectives of tacit knowledge: the \textit{element} where tacit knowledge is revealed or manifested in graphic design, the \textit{action} performed by the subject to reveal their tacit knowledge, and the \textit{purpose} of using the knowledge. In prior work on the tacit knowledge of design, Bernal et al. ~\cite{bernal2015role} investigated whether tacit knowledge could be supported by mapping the actions of design experts (i.e., Generation, Evaluation, Selection, and Integration) to a computational approach that can support these actions. However, in order to define tacit knowledge as knowledge that can be captured and reused, we need to 1) concretize abstract actions, such as \textit{Integration} ~\cite{bernal2015role}, into low-level actions, and 2) specify the target or element in which this action is performed. Furthermore, the context in which the knowledge is used, including the background and purpose of the knowledge, needs to be specified when the designer applies this knowledge. Therefore, we defined not only the specific \textit{actions} of the designer, but also the \textit{element} and \textit{purpose} in which these actions are performed as a scheme to code and analyze instances.

\subsubsection{Open Coding Process}
Two authors conducted an open coding ~\cite{corbin1990grounded, khandkar2009open} on the collected tacit knowledge instances from the participants (Total: 123; Mean: 12.3; SD: 1.49). Before starting the coding process, the authors excluded instances where two or more schemes did not emerge. For example, the instance \textit{``Design an appropriate action to achieve a certain purpose''} (P2) was excluded. A total of 10 instances were excluded, and coding was carried out for 123 instances as follows. For example, in the \textit{``Adjusting the right font size, boldness, line spacing, and margin according to the hierarchy of the content''} (P10), the element would be \textit{Font}, the action is \textit{Adjust}, and the purpose is \textit{Considering the importance of information}. The two authors assigned codes only when these were specifically stated in the instance for element, action, and purpose.

There were instances without a specified purpose (n=10) and instances that could be revealed through all elements (n=22) in the collected data points. For example, \textit{``The ability to find the right amount of blank space in design’’} (P7) doesn't indicate the purpose of this knowledge. Here, the authors did not make additional interpretations but only performed coding for the scheme specified in the instance (in this case, element and action). As another example, the \textit{``The ability to design well by blending in, rather than just following the trend''} (P2) includes a purpose to \textit{In line with design trend}, but since it does not specifically state through which element it is, the authors have assigned the element of this instance as an \textit{Overall} code. The meaning of the \textit{Overall} code implies that this knowledge spans all elements.

As a result of iterative qualitative coding for each of the \textit{element}, \textit{action}, and \textit{purpose} of the collected 123 tacit knowledge instances, we defined 4 categories and 13 codes (exclude \textit{Overall}) in the element, 3 categories and 11 codes in action, and 5 categories and 16 codes in purpose (\autoref{tab:instance_results}).
\label{section_opencoding}

\begin{table*}[!ht]
\caption{Tacit knowledge in graphic design. The results of the open coding process are in Section~\ref{section_opencoding}. The table shows the defined code and category. The relatively lowest/highest count of the category (in each coding scheme) is highlighted in {\color{red} red}/{\color{teal} green}.}
\Description{Table 3 shows the code and category of tacit knowledge in graphic design. Since our qualitative coding was conducted regarding the instance’s element, action, and purpose, three coding schemes are on the left side. Then, the category, code, code count, and category count are shown. There are five categories (inner design element, design work pattern, contents, outer design source, and overall) in element, three categories (cognition, utilization, and manipulation) in action, and seven categories (environment, audience, visual, individual design style, design contents, and empty) in purpose. The highest number of categories is colored green, and the lowest number of categories is colored red.
}
\label{tab:instance_results}
\resizebox{\linewidth}{!}{
\makegapedcells
\def\arraystretch{1.2}%
\begin{tabular}{lllcc}
    \cmidrule[1.5pt]{1-5}
    \makecell[ll]{\textbf{Coding}\\\textbf{scheme}} & \textbf{Category} & \textbf{Code (example of instance)} & \makecell[cc]{\textbf{Code}\\count} & \makecell[cc]{\textbf{Category}\\count (\%)} \\
    \cmidrule[1.5pt]{1-5}

    \multirow{14}{*}{Element} & \multirow{7}{*}{\makecell[ll]{Inner design\\element}} & \textbf{Color} (e.g., Pairing \textbf{colors} based on what the target demographic favors by P9) & 15 & {\color{teal} \multirow{7}{*}{\textbf{64 (56.64\%})}}\\
    \multirow{14}{*}{} & \multirow{7}{*}{} & \textbf{Font} (e.g., Combining aesthetically pleasing \textbf{fonts} by P2) & 12 & \multirow{7}{*}{}\\
    \multirow{14}{*}{} & \multirow{7}{*}{} & \textbf{Image} (e.g., Harmoniously arranging \textbf{images} to match the background by P2) & 3 & \multirow{7}{*}{}\\
    \multirow{14}{*}{} & \multirow{7}{*}{} & \textbf{Layout} (e.g., Creating a \textbf{layout} that guides the audience's actions by P4) & 6 & \multirow{7}{*}{}\\
    \multirow{14}{*}{} & \multirow{7}{*}{} & \textbf{Size/ratio} (e.g., A sense knowledge of proper \textbf{size and proportion} for balanced design by P1) & 8 & \multirow{7}{*}{}\\
        \multirow{14}{*}{} & \multirow{7}{*}{} & \textbf{Space} (e.g., Judging visually comfortable \textbf{margin space} by P6) & 7 & \multirow{7}{*}{}\\
    \multirow{14}{*}{} & \multirow{7}{*}{} & \textbf{User experience} (e.g., Predicting the mental model of the \textbf{target audience} by P5) & 13 & \multirow{7}{*}{}\\
    \cline{2-5}
    \multirow{14}{*}{}& \multirow{2}{*}{\makecell[ll]{Design work\\pattern}} & \textbf{Design tool} (e.g., Knowing how to use various \textbf{tools} according to the design situation by P8) & 2 & \multirow{2}{*}{8 (7.08\%)}\\
    \multirow{14}{*}{}& \multirow{2}{*}{} & \textbf{Work style} (e.g., \textbf{A work pattern} conducting sufficient design research by P10) & 6 & \multirow{2}{*}{}\\
    \cline{2-5}
    \multirow{14}{*}{}& \multirow{2}{*}{Contents} & \textbf{Information hierarchy} (e.g., Judging the importance of \textbf{information} based on the content by P9)& 3 & {\color{red} \multirow{2}{*}{5 (4.42\%)}}\\
    \multirow{14}{*}{}& \multirow{2}{*}{} & \textbf{Writing} (e.g., Understanding subtle differences in nuance in \textbf{text phrases} by P5) & 2 & \multirow{2}{*}{}\\
    \cline{2-5}
    \multirow{14}{*}{}& \multirow{2}{*}{\makecell[ll]{Outer\\design source}} & \textbf{Design theory} (e.g., Arranging the layout based on the \textbf{golden ratio rule} by P6) & 6 & \multirow{2}{*}{11 (9.73\%)}\\
    \multirow{14}{*}{}& \multirow{2}{*}{} & \textbf{Reference} (e.g., Feeling inspired by unrelated \textbf{sources} by P6) & 5 & \multirow{2}{*}{}\\
    \cline{2-5}
    \multirow{14}{*}{}& \multirow{1}{*}{Overall} & \textbf{Overall} (e.g., Adapting styles that are in line with the current trends by P5 -> ``Various elements affect the style'')& 25 &\multirow{1}{*}{25 (22.12\%)}\\
    \cmidrule[1.5pt]{1-5}
    
    \multirow{11}{*}{Action} & \multirow{4}{*}{Cognition} & \textbf{Possess} (e.g., \textbf{Possessing} knowledge about the usage of easily readable fonts by P10) & 10 & \multirow{4}{*}{37 (32.74\%)}\\
    \multirow{11}{*}{} & \multirow{4}{*}{} & \textbf{Feel} (e.g., \textbf{Feeling} inspired by unrelated sources by P1) & 8 & \multirow{4}{*}{}\\
    \multirow{11}{*}{} & \multirow{4}{*}{} & \textbf{Predict} (e.g., \textbf{Predicting} design error cases that others may not consider by P6) & 9 & \multirow{4}{*}{}\\
    \multirow{11}{*}{} & \multirow{4}{*}{} & \textbf{Judge} (e.g., \textbf{Judging} the harmony between text and image by P8) & 10 & \multirow{4}{*}{}\\
    \cline{2-5}
    \multirow{11}{*}{}& \multirow{2}{*}{Utilization} & \textbf{Find} (e.g., \textbf{Finding} fonts that fit well with the style/mood by P8) & 3 & {\color{red} \multirow{2}{*}{15 (13.27\%)}}\\
    \multirow{11}{*}{}& \multirow{2}{*}{} & \textbf{Adapt} (e.g., Appropriately \textbf{adapting} basic design principles according to the design situation by P7) & 12 & \multirow{2}{*}{}\\
    \cline{2-5}
    \multirow{11}{*}{}& \multirow{5}{*}{Manipulation} & \textbf{Create} (e.g., \textbf{Creating} a layout that guides the user's actions by P4) & 23 & {\color{teal} \multirow{5}{*}{\textbf{61 (53.98\%)}}}\\
    \multirow{11}{*}{}& \multirow{5}{*}{} & \textbf{Arrange} (e.g., \textbf{Arranging} objects so that the space doesn't look empty by P9) & 6 & \multirow{5}{*}{}\\
    \multirow{11}{*}{} & \multirow{5}{*}{} & \textbf{Adjust} (e.g., \textbf{Adjusting} the intensity of visualization to balance the design elements by P2) & 17 & \multirow{5}{*}{}\\
    \multirow{11}{*}{} & \multirow{5}{*}{} & \textbf{Combine} (e.g., \textbf{Combining} aesthetically pleasing colors by P2)  & 8 & \multirow{5}{*}{}\\
    \multirow{11}{*}{} & \multirow{5}{*}{} & \textbf{Define} (e.g., \textbf{Defining} the proper size of an element by P1) & 7 & \multirow{5}{*}{}\\
    \cmidrule[1.5pt]{1-5}
    
    \multirow{20}{*}{Purpose} & \multirow{3}{*}{Environment} & \textbf{External constraint} (e.g., Possessing the knowledge of creating designs \textbf{at a low cost} by P5) & 14 & \multirow{3}{*}{24 (21.24\%)}\\
    \multirow{20}{*}{} & \multirow{3}{*}{} & \textbf{Design requirement} (e.g., Adapting design elements \textbf{to conform to the brand's identity} by P7). & 7 & \multirow{3}{*}{}\\
    \multirow{20}{*}{} & \multirow{3}{*}{} & \textbf{In line with trends} (e.g., Adapting the design style \textbf{to match a trend} by P5) & 3 & \multirow{3}{*}{}\\
    \cline{2-5}
    
    \multirow{20}{*}{}& \multirow{5}{*}{Audience} & \textbf{Favored by the public} (e.g., Creating designs with a \textbf{focus on mainstream appeal} by P3) & 2 & \multirow{5}{*}{32 (28.32\%)}\\
    \multirow{20}{*}{}& \multirow{5}{*}{} & \textbf{User-friendly} (e.g., Predicting the design \textbf{from the user's perspective} by P7) & 7 & \multirow{5}{*}{}\\
    \multirow{20}{*}{}& \multirow{5}{*}{} & \textbf{Reducing errors} (e.g., Predicting design \textbf{error cases} that others may not consider by P6) & 7 & \multirow{5}{*}{}\\
    \multirow{20}{*}{}& \multirow{5}{*}{} & \textbf{Considering the target audience} (e.g., Pairing \textbf{colors} based on what the target \textbf{demographic favors} by P9) & 7 & \multirow{5}{*}{}\\
    \multirow{20}{*}{}& \multirow{5}{*}{} & \textbf{Guiding the audience} (e.g., Determining the proper size of an element \textbf{to guide the user's attention} by P9) & 9 & \multirow{5}{*}{}\\
    \cline{2-5}
    
    \multirow{20}{*}{}& \multirow{5}{*}{Visual} & \textbf{Enhancing readability} (e.g., \textbf{Increasing readability} by adjusting the attributes of the font by P2) & 7 & {\color{teal} \multirow{5}{*}{\textbf{34 (30.09\%})}}\\
    \multirow{20}{*}{}& \multirow{5}{*}{} & \textbf{Balanced} (e.g., Arranging objects so that the space \textbf{doesn't look empty} by P9) & 5 & \multirow{5}{*}{}\\
    \multirow{20}{*}{}& \multirow{5}{*}{} & \textbf{Aesthetically pleasing} (e.g., Combining \textbf{aesthetically pleasing} colors by P2) & 13 & \multirow{5}{*}{}\\
    \multirow{20}{*}{}& \multirow{5}{*}{} & \textbf{Unexpected} (e.g., Finding \textbf{unexpected} color palette from completely different domains by P7) & 4 & \multirow{5}{*}{}\\
    \multirow{20}{*}{}& \multirow{5}{*}{} & \textbf{Harmonious} (e.g., Judging the \textbf{harmony} between text and images by P8) & 5 & \multirow{5}{*}{}\\
    \cline{2-5}
    
    \multirow{20}{*}{}& \multirow{2}{*}{\makecell[ll]{Individual\\design style}} &  \multirow{2}{*}{\textbf{Preference} (e.g., Possessing a \textbf{personal style preference} by P8)} & \multirow{2}{*}{3} &{\color{red} \multirow{2}{*}{3 (2.65\%)}}\\
    \multirow{20}{*}{}& \multirow{2}{*}{} & \multirow{2}{*}{} &  & \multirow{2}{*}{}\\
    \cline{2-5}
    
    \multirow{20}{*}{}& \multirow{4}{*}{\makecell[ll]{Design\\contents}} & \multirow{2}{*}{\makecell[ll]{\textbf{Considering the importance of information}\\(e.g., Adjusting the font's emphasis according to the \textbf{information importance} by P2)}} & \multirow{2}{*}{7} & \multirow{4}{*}{12 10.62\%)}\\
    \multirow{20}{*}{}& \multirow{4}{*}{} & \multirow{2}{*}{} & \multirow{2}{*}{} & \multirow{4}{*}{}\\
    \multirow{20}{*}{}& \multirow{4}{*}{} & \multirow{2}{*}{\makecell[ll]{\textbf{In line with the design theme}\\(e.g., Defining \textbf{what part of the topic to focus on} when setting a design concept by P10)}} & \multirow{2}{*}{5}& \multirow{4}{*}{}\\
    \multirow{20}{*}{}& \multirow{4}{*}{} & \multirow{2}{*}{} & \multirow{2}{*}{} & \multirow{4}{*}{}\\
    \cline{2-5}
    
    \multirow{20}{*}{}& \multirow{1}{*}{Empty} & \textbf{Empty} (e.g., Defining an element's size by P1 -> ``There is no explicitly described purpose'') & 8 & \multirow{1}{*}{8 (7.08\%)}\\
    \cmidrule[1.5pt]{1-5}
\end{tabular}
}
\end{table*}

\subsubsection{Analysis on Characteristics of Instances}
To specify the characteristics of the different elements, actions, and purposes of tacit knowledge from the category level of instance, we analyzed the interviewees' annotations from the study. We calculated how often each characteristic was annotated as present in each tacit knowledge instance. Through this analysis, we report how the distribution of tacit knowledge characteristics exists and what differences exist in each category.
\label{section_characteristicsanalysismethod}

\begin{figure*}[!ht]
  \centering
  \includegraphics[width=0.9\linewidth]{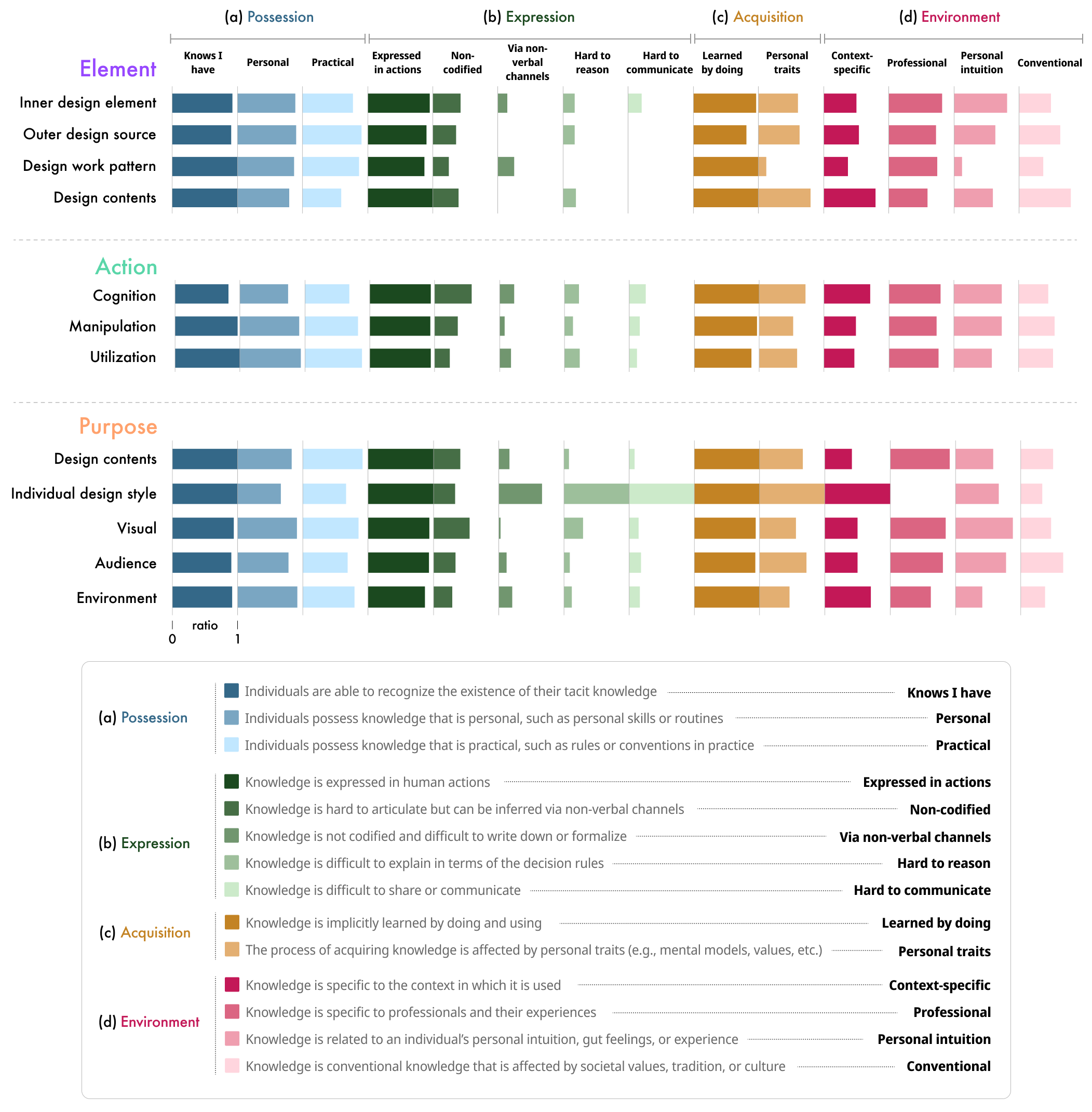}
  \caption{Annotation analysis results of tacit knowledge characteristics. The chart shows how many annotations are obtained from the 10 graphic designers in each category of element, action, and purpose as a ratio (the calculation method is in Section 
  \ref{section_characteristicsanalysismethod}. The legend of (a) to (d) indicates each category of ~\autoref{tab:tacit-characteristics-list}.}
  \Description{Figure 1. Figure 1 shows that the annotation analysis results in tacit knowledge instances and characteristics. The analysis was also conducted regarding the three coding schemes (element, action, and purpose). The bar chart is illustrated, and the ratio of each characteristic annotated is shown in each category level. The below part contains the legend of characteristics that came from Table 1. Each characteristic has a different color.
}
  \label{fig:Characteristics}
\end{figure*}
\subsection{Findings}

\subsubsection{RQ1-1: Elements, Actions, and Purpose of Tacit Knowledge in Graphic Design}
Our analysis results show that graphic designers use tacit knowledge in elements related to \textit{inner-design elements}, through actions centered on \textit{manipulation}, and to create \textit{visually pleasing} designs at the category level (\autoref{tab:instance_results}-Category). 

The most salient elements where tacit knowledge is evident are the \textit{color}, \textit{font}, and \textit{user experience} at the code level (\autoref{tab:instance_results}-Code). Among the actions, graphic designers primarily engage in \textit{creation} and \textit{manipulation}-related actions. With respect to \textit{purpose}, their tacit knowledge is used to satisfy \textit{external design constraints} (e.g., a limited budget or time) while striving to create \textit{visually appealing} designs. Additionally, a significant portion of tacit knowledge's purpose is to consider the actual \textit{audience} who will face the design.
\label{RQ1-1}

\subsubsection{RQ1-2: Element, Where Tacit Knowledge is Revealed or Manifested}
The coding results show that the categories of \textit{elements} include those \textit{inner design elements}, \textit{design work pattern}, \textit{contents}, and \textit{outer design sources} (\autoref{tab:instance_results}-Element's Category). Regarding the characteristics of tacit knowledge, the instances revealed through inner design elements require the most intuition and simultaneously have characteristics of professional knowledge (\autoref{fig:Characteristics}-d). 

On the other hand, the \textit{work pattern}-related element's instances are the least intuitive and show the easiest characteristics in reasoning and communication. \textit{``The ability to design various design proposals quickly and in large quantities''} mentioned by P7 is one of the instances revealed from the work pattern. Regarding acquiring tacit knowledge, graphic designers mostly learn while doing design work, regardless of the type of element (\autoref{fig:Characteristics}-c). Tacit knowledge instances appearing as \textit{inner design elements} are somewhat expressible through actions, but they show the most difficult characteristics in terms of codification and communication (\autoref{fig:Characteristics}-b). P1, P3, and P9 have stated that fine adjustments of design elements can lead to a good design. They also mentioned that while this can be discovered during the design modification process, they often explain the rationale of modification to others using abstract terms.
\label{RQ1-2}

\subsubsection{RQ1-3: Action, Designer's Actions to Practice Tacit Knowledge}
Graphic designers' actions to use tacit knowledge can be largely categorized into \textit{cognition}, \textit{utilization}, and \textit{manipulation} aspects (\autoref{tab:instance_results}-Action's Category). \textit{Cognition} includes actions such as possess or feel. The \textit{utilization} contains tacit knowledge that can be performed through finding and adapting. The \textit{manipulation} category has tacit knowledge that can be represented through actions such as creating or arranging.

From the perspective of characteristics, it is difficult to know that the \textit{cognitive-related actions} of tacit knowledge are owned compared to other action categories (\autoref{fig:Characteristics}-a). It also shows that it is the most difficult to codify and communicate. Instances of tacit knowledge that can fall into \textit{cognitive-related actions} category include \textit{``Possessing knowledge about the usage of easily readable fonts''}(P10). In the interview, several designers (P3-P5, P7, and P10) explained that making judgments or predictions about certain design problems or situations is difficult to explain verbally and appears differently among experts. They also mentioned that tacit knowledge used by performing such actions highly depends on personal intuition. The most dominant actions to use tacit knowledge are in the \textit{manipulation} category ~\autoref{tab:instance_results}. Specifically, \textit{creation-related action} shows the characteristics of being most influenced by personal intuition and being conventional (\autoref{fig:Characteristics}-c and d).

Relatively, the smallest actions are the \textit{utilization-related actions}, which include find and adapt (\autoref{tab:instance_results}). Instances associated with these actions are \textit{``Finding fonts that fit well with the style/mood''} (P8) and \textit{``Appropriately adapting basic design principles according to the design situation''} (P7). In ~\autoref{fig:Characteristics}, \textit{utilization} shows the characteristics of difficulty in communication and reasoning.
\label{RQ1-3}

\subsubsection{RQ1-4: Purpose, Goal of Tacit Knowledge}
Tacit knowledge in graphic design is used for five main purposes: \textit{Individual design style}, \textit{environment}, \textit{visual}, \textit{design content}, and \textit{audience-related purpose} (\autoref{tab:instance_results}-Purpose's Category). Tacit knowledge is mostly used for \textit{visual purposes}, with the least associated with \textit{individual design style}. Also, \textit{audience-} and \textit{environment-related purposes} were shown as one of the main purposes of tacit knowledge in graphic design.

The tacit knowledge instances with the most difficult codifying characteristic are those performed for a visual-related purpose (\autoref{fig:Characteristics}-b). The instances aim to be aesthetically pleasing, like `\textit{`Combining aesthetically pleasing colors''} (P2), or \textit{``Creating visually comfortable margins''} (P6). P4 and P7 said that a visually attractive graphic design is hard to explain, but a designer with this knowledge reveals it through their editing or creating actions. The most intuitive tacit knowledge instances are included here, and reasoning is also rather hard (\autoref{fig:Characteristics}-b and d).

Audience is a group that graphic designers always consider essential (P1-P10), and tacit knowledge with a related purpose has the most conventional characteristics and requires relatively high intuition (\autoref{fig:Characteristics}). This includes the instances of \textit{``Creating a layout that guides the audience's action''} (P4). Tacit knowledge instances with an environment-related purpose include instances of \textit{''Possessing the knowledge of creating designs at a low cost''} (P5). The instances with this purpose show relatively high context-specific characteristics, but it has the characteristic least influenced by personal traits and intuition (\autoref{fig:Characteristics}-c and d).

Instances with a purpose related to \textit{individual style} are the least frequent (\autoref{tab:instance_results}). However, among all categories of tacit knowledge (\autoref{tab:instance_results}), instances are the most difficult to reason and communicate, and they can be expressed through non-verbal channels (\autoref{fig:Characteristics}-b). An example of an instance including this category is \textit{''Possessing a personal style preference''} (P8). P3 and P8 commented that personal style or preference could make the design communication difficult, but this is not tacit knowledge only for the professional.
\label{RQ1-4}

%% file: sections/6_systematic-literature-review.tex
\section{Systematic Literature Review: Exploring Graphic Design Approaches}

To investigate gaps between the investigated instances and the prior approaches that support the graphic design process (e.g., design creation, knowledge sharing, etc.), we conduct a systematic literature review (SLR) on approaches, encompassing systems, techniques, and methodologies related to graphic design. By annotating the elements, actions, and purposes of instances supported explicitly by the approach, we investigate how existing approaches address different types of tacit knowledge in graphic design. This SLR aims to uncover what categories or codes of tacit knowledge instances remain unsupported or under-supported and to analyze the reasons through the characteristics of tacit knowledge in section ~\ref{section_characteristic}, answering these research questions: \label{section_rq2}

\begin{itemize}
    \item RQ2: How do prior \textbf{approaches} in graphic design support the usage of tacit knowledge, and what parts lack support?
    \begin{itemize}
        \item RQ2-1: Which \textit{elements} of tacit knowledge get adequate or insufficient coverage by prior approaches?
        \item RQ2-2: Which \textit{actions} of tacit knowledge get adequate or insufficient coverage by prior approaches?
        \item RQ2-3: Which \textit{purposes} of tacit knowledge get adequate or insufficient coverage by prior approaches?
    \end{itemize}
\end{itemize}

\subsection{Sampling and Annotation}
\subsubsection{Search Keyword and Exclusion Criteria}
We first searched for literature with broader keywords in the ACM Digital Library to identify existing approaches without constraint for publication year. The keywords were ``design communication'', ``design knowledge'', ``design expression'', ``design rationale'', and ``graphic design''. Then, we narrowed our search to papers suggesting artifact contributions (e.g., interfaces, algorithms) or those presenting frameworks or methods, that could be implemented as tools, to support graphic designers. Accordingly, empirical studies (e.g., ~\cite{herring2009getting}) found in the sampling process were excluded from the review. 

With this exclusion criterion, we collected 108 papers using the keywords. Then, we filtered these papers based on their relatedness to the graphic design domain. For example, we included the papers if they clearly explain their focus on the graphic design domain. However, we excluded papers that focused on domains unrelated to graphic design, such as architecture, fine art, or painting. We filtered out 65 papers and arrived at a final set of 43 papers. To explore diverse approaches, we performed an exploratory search of papers (19 papers) meeting the criteria but not identified in the initial search, as well as papers citing the previously discovered papers. Consequently, we examined a total of 62 papers.

\subsubsection{Annotation Process}
Two authors conducted the annotation process. An approach was considered to cover a knowledge instance if it explicitly detailed the specific elements, actions, and purpose of instances in their approach. Both authors primarily reviewed the introduction and methodology sections of a sampled paper, discussing and annotating the covered elements, actions, and purpose (\autoref{tab:instance_results}). In the case of conflicts during the annotation, the authors engaged in iterative discussions, establishing rules for conflicting annotations to prevent further conflicts in subsequent annotations. Discussions continued until the authors reached a consensus and completed the entire annotation process.

The annotation process is illustrated with the following example: In the methodology section, if the approach mentions element codes and indicates system support for manipulation within those elements, the paper is marked as addressing them. As an example, Vinci ~\cite{guo2021vinci}, a design system supporting the process of creating advertisement posters, takes reference images and the desired concept of the poster to adjust elements such as color, font, image, and layout, ultimately providing a completed poster. In this case, we annotated Vinci's handled elements as \textit{color}, \textit{font}, \textit{image}, \textit{layout}, and \textit{reference}. The annotation was also performed for the actions explicitly mentioned in the paper. Returning to the example of Vinci, this approach automatically generates designs on behalf of the designer, and the designer also arranges and adjusts the generated design. We annotated Vinci's supported actions as \textit{create}, \textit{arrange}, and \textit{adjust}. Regarding the purpose of an approach, we found their aims primarily in the introduction section. Vinci aims to create visually appealing and readable posters within the given design requirements, so we annotated the purpose as \textit{aesthetically pleasing}, \textit{enhancing readability}, and \textit{satisfying design requirements} codes in ~\autoref{tab:instance_results}.

\subsection{Analysis Method}
To analyze the annotation result, the total amount of \textbf{O} marks was counted in each element, action, and purpose code level (\autoref{tab:instance_results}-Code). We divided this by the total number of reviewed papers to calculate the percentage of coverage within the sampled papers in each code level. Then, we averaged each code's coverage value in a category to get the average coverage values in each category (\textit{Approach avg ratio} in ~\autoref{tab:slr_results}). We compared this ratio with ~\autoref{tab:instance_results}'s category ratio to see which types of categories are relatively less covered from prior approaches. This method reveals the differences between the distribution of tacit knowledge instances and those covered by the approaches. Additionally, drawing from the findings in Sections ~\ref{RQ1-2} to ~\ref{RQ1-4}, we examine the reasons behind the limited coverage of specific instances by describing the characteristics of categories.

\subsection{Findings}
\label{section_slr_findings}

\begin{table*}[!ht]
\caption{Systematic literature review result. This table shows the SLR results. We annotated 62 prior approaches for graphic design to the code of ~\autoref{tab:instance_results} and calculated the ratio of coverage in each code. This table shows the average ratio of each code at the category level. The relatively lowest/highest coverage ratio of the category (in each coding scheme) is highlighted in {\color{red} red}/{\color{teal} green}. The rightest side shows the annotated approaches.}
\Description{Table 4: Table 4 shows the annotation results of our systematic literature review. The qualitative coding results are namely as shown in Table 3 same. The right side columns show the prior approaches that were annotated to this code with the average ratio of approaches in the category level. The highest number of categories is colored green, and the lowest number of categories is colored red.}
\label{tab:slr_results}
\resizebox{\linewidth}{!}{
\def\arraystretch{1.2}%
\begin{tabular}{lllcp{\columnwidth}}
    \cmidrule[1.5pt]{1-5}
    \textbf{\makecell[ll]{Coding\\scheme}} & \textbf{Category} & \textbf{Code} & \makecell[ll]{\textbf{Approach}\\\textbf{avg ratio (\%)}} & \textbf{Approaches}\\
    \cmidrule[1.5pt]{1-5}
    \multirow{14}{*}{\textbf{Element}} & \multirow{7}{*}{Inner design element} & Color & \multirow{7}{*}{21.66\%} &\cite{chen2021cocapture,wang2020learning,guo2021vinci,newman2000sitemaps,xie2021canvasemb,chilton2019visiblends,zhao2021selective,zhao2018characterizes,yang2019artistic,chilton2021visifit,kim2022stylette,tanner2019poirot,kumar2011bricolage,ritchie2011d} \\
    \multirow{14}{*}{} & \multirow{7}{*}{} & Font &\multirow{7}{*}{}&\cite{guo2021vinci,rebelo2021exploring,newman2000sitemaps,ross2020wordblender,xie2021canvasemb,chilton2019visiblends,zhao2018characterizes,yang2019artistic,choi2018fontmatcher,lopes2020adea,kim2022stylette,tanner2019poirot,kumar2011bricolage,ritchie2011d} \\
    \multirow{14}{*}{} & \multirow{7}{*}{} & Image &\multirow{7}{*}{} &\cite{wang2020learning,guo2021vinci,ross2020wordblender,xie2021canvasemb,chilton2019visiblends,zhao2021selective,zhao2018characterizes,yang2019artistic,chilton2021visifit,kumar2011bricolage,ritchie2011d} \\
    \multirow{14}{*}{} & \multirow{7}{*}{} & Layout &\multirow{7}{*}{}&\cite{chen2021cocapture,zheng2019content,wang2020learning,roth1994interactive,guo2021vinci,rebelo2021exploring,maudet2017beyond,kim1993providing,xie2021canvasemb,ueno2021continuous,kikuchi2021constrained,yang2019artistic,mackinlay1986automating,o2015designscape,liu2018data,dayama2020grids,zhao2020iconate,swearngin2020scout,kumar2011bricolage,ritchie2011d} \\
    \multirow{14}{*}{} & \multirow{7}{*}{} & Size/ratio &\multirow{7}{*}{}&\cite{chen2021cocapture,zheng2019content,wang2020learning,guo2021vinci,rebelo2021exploring,maudet2017beyond,kim1993providing,ross2020wordblender,xie2021canvasemb,ueno2021continuous,zhao2018characterizes,kikuchi2021constrained,yang2019artistic,chilton2021visifit,mackinlay1986automating,o2015designscape,liu2018data,dayama2020grids,kim2022stylette,swearngin2020scout,tanner2019poirot,kumar2011bricolage,ritchie2011d}\\
    \multirow{14}{*}{} & \multirow{7}{*}{} & Space &\multirow{7}{*}{}&\cite{zheng2019content,guo2021vinci,kim1993providing,xie2021canvasemb,ueno2021continuous,kim2022stylette,swearngin2020scout,tanner2019poirot,kumar2011bricolage}\\
    \multirow{14}{*}{} & \multirow{7}{*}{} & User experience &\multirow{7}{*}{} &\cite{newman2000sitemaps,kusano2013scenario,tan2012development}\\
    \cline{2-5}
    
    \multirow{14}{*}{}& \multirow{2}{*}{Design work pattern} & Design tool &{\color{teal} \multirow{2}{*}{44.35\%}} &\cite{chen2021cocapture,haruno2016modern,milota2004modality,roth1994interactive,lewien2021gazehelp,klemmer2002web,newman2000sitemaps,kim1993providing,o2018charrette,chen2019gallery,kim2022stylette,swearngin2020scout,fogarty2008cueflik,tanner2019poirot,lee2020guicomp} \\
    \multirow{14}{*}{}& \multirow{2}{*}{} & Work style &\multirow{2}{*}{}&\cite{li2012sketchcomm,chen2021cocapture,wang2020learning,roth1994interactive,guo2021vinci,klemmer2002web,maudet2017beyond,newman2000sitemaps,kim1993providing,soria2019collecting,gutierrez2018re,chilton2019visiblends,ueno2021continuous,o2018charrette,rasmussen2019co,mackinlay1986automating,date2018paperprinting,o2015designscape,kang2021metamap,liu2018data,lowgren1992knowledge,lekschas2021ask,tan2012development,suleri2019ui,yen2020decipher,ngoon2021shown,dayama2020grids,kang2018paragon,swearngin2020scout,fogarty2008cueflik,foong2021crowdfolio,lee2020guicomp,huang2019swire,zilouchian2015procid,kim2021winder,koch2020semanticcollage,bunian2021vins,mozaffari2022ganspiration,kumar2011bricolage,ritchie2011d}\\
    \cline{2-5}

    \multirow{14}{*}{}& \multirow{2}{*}{Contents} & Information hierarchy &{\color{red} \multirow{2}{*}{2.42\%}}&\cite{rebelo2021exploring,xie2021canvasemb,kumar2011bricolage}\\
    \multirow{14}{*}{}& \multirow{2}{*}{} & Writing &\multirow{2}{*}{}&[]\\
    \cline{2-5}
    
    \multirow{14}{*}{}& \multirow{2}{*}{Outer design source} & Design theory &\multirow{2}{*}{25.00\%} &\cite{maudet2017beyond,kim1993providing,lowgren1992knowledge,suleri2019ui,kang2018paragon}\\
    \multirow{14}{*}{}& \multirow{2}{*}{} & Reference &\multirow{2}{*}{}&\cite{li2012sketchcomm,wang2020learning,guo2021vinci,maudet2017beyond,ross2020wordblender,gutierrez2018re,xie2021canvasemb,chilton2019visiblends,ueno2021continuous,lin2021learning,chen2019gallery,yang2019artistic,mackinlay1986automating,kang2021metamap,suleri2019ui,zhao2020iconate,kang2018paragon,kim2022stylette,fogarty2008cueflik,lee2020guicomp,huang2019swire,koch2020semanticcollage,bunian2021vins,mozaffari2022ganspiration,kumar2011bricolage,ritchie2011d}\\
    \cline{2-5}

    \multirow{14}{*}{}& Overall element & Overall & N/A & N/A\\
    \cmidrule[1.0pt]{1-5}

    \multirow{11}{*}{\textbf{Action}} & \multirow{4}{*}{Cognition} & \textbf{Possess} &{\color{red} \multirow{4}{*}{13.71\%}}&\cite{lewien2021gazehelp,klemmer2002web,soria2019collecting,gutierrez2018re,o2018charrette,rasmussen2019co,tan2012development,kang2018paragon,foong2021crowdfolio,lee2020guicomp,zilouchian2015procid,ritchie2011d}\\
    \multirow{11}{*}{} & \multirow{4}{*}{} & Feel &\multirow{4}{*}{}&\cite{kusano2013scenario} \\
    \multirow{11}{*}{} & \multirow{4}{*}{} & Predict &\multirow{4}{*}{}&\cite{chen2021cocapture,newman2000sitemaps,zhao2018characterizes,date2018paperprinting,bylinskii2017learning,lekschas2021ask,tan2012development,yen2020decipher,kang2018paragon,foong2021crowdfolio,lee2020guicomp}\\
    \multirow{11}{*}{} & \multirow{4}{*}{} & Judge &\multirow{4}{*}{}&\cite{o2018charrette,date2018paperprinting,lowgren1992knowledge,lekschas2021ask,yen2020decipher,kang2018paragon,foong2021crowdfolio,lee2020guicomp,zilouchian2015procid,kim2021winder}\\
    \cline{2-5}
    
    \multirow{11}{*}{}& \multirow{2}{*}{Utilization} & Find &{\color{teal} \multirow{2}{*}{31.45\%}}&\cite{klemmer2002web,kim1993providing,soria2019collecting,ross2020wordblender,gutierrez2018re,xie2021canvasemb,rasmussen2019co,chen2019gallery,kang2021metamap,choi2018fontmatcher,tan2012development,suleri2019ui,yen2020decipher,ngoon2021shown,kang2018paragon,kim2022stylette,fogarty2008cueflik,tanner2019poirot,lee2020guicomp,huang2019swire,zilouchian2015procid,koch2020semanticcollage,bunian2021vins,mozaffari2022ganspiration,ritchie2011d}\\
    \multirow{11}{*}{}& \multirow{2}{*}{} & Adapt &\multirow{2}{*}{}&\cite{li2012sketchcomm,chen2021cocapture,zheng2019content,wang2020learning,roth1994interactive,maudet2017beyond,kim1993providing,chilton2019visiblends,yang2019artistic,ngoon2021shown,zhao2020iconate,kang2018paragon,tanner2019poirot,kumar2011bricolage}\\
    \cline{2-5}

    \multirow{11}{*}{}& \multirow{5}{*}{Manipulation} & Create &\multirow{5}{*}{23.23\%}&\cite{haruno2016modern,zheng2019content,wang2020learning,guo2021vinci,rebelo2021exploring,maudet2017beyond,kim1993providing,ross2020wordblender,xie2021canvasemb,chilton2019visiblends,ueno2021continuous,kikuchi2021constrained,yang2019artistic,kusano2013scenario,chilton2021visifit,mackinlay1986automating,liu2018data,lopes2020adea,dayama2020grids,swearngin2020scout,mozaffari2022ganspiration}\\
    \multirow{11}{*}{}& \multirow{5}{*}{} & Arrange &\multirow{5}{*}{} &\cite{zheng2019content,guo2021vinci,rebelo2021exploring,kim1993providing,chilton2019visiblends,kikuchi2021constrained,yang2019artistic,mackinlay1986automating,liu2018data,dayama2020grids,swearngin2020scout}\\
    \multirow{11}{*}{} & \multirow{5}{*}{} & Adjust &\multirow{5}{*}{} &\cite{li2012sketchcomm,chen2021cocapture,zheng2019content,wang2020learning,guo2021vinci,rebelo2021exploring,maudet2017beyond,newman2000sitemaps,kim1993providing,ross2020wordblender,chilton2019visiblends,ueno2021continuous,zhao2021selective,zhao2018characterizes,yang2019artistic,kusano2013scenario,chilton2021visifit,mackinlay1986automating,o2015designscape,liu2018data,choi2018fontmatcher,dayama2020grids,kim2022stylette,swearngin2020scout,tanner2019poirot,kumar2011bricolage}\\
    \multirow{11}{*}{} & \multirow{5}{*}{} & Combine &\multirow{5}{*}{} &\cite{wang2020learning,ross2020wordblender,chilton2019visiblends,yang2019artistic,chilton2021visifit,zhao2020iconate,kumar2011bricolage} \\
    \multirow{11}{*}{} & \multirow{5}{*}{} & Define &\multirow{5}{*}{} &\cite{maudet2017beyond,kim1993providing,lin2021learning,o2015designscape,tan2012development,fogarty2008cueflik,koch2020semanticcollage} \\
    \cmidrule[1.0pt]{1-5}

    \multirow{16}{*}{purpose} & \multirow{3}{*}{Environment} & Outer constraint &{\color{teal} \multirow{3}{*}{29.03\%}}&\cite{li2012sketchcomm,haruno2016modern,wang2020learning,roth1994interactive,guo2021vinci,maudet2017beyond,newman2000sitemaps,gutierrez2018re,chilton2019visiblends,ueno2021continuous,o2018charrette,rasmussen2019co,kikuchi2021constrained,chen2019gallery,yang2019artistic,kusano2013scenario,chilton2021visifit,mackinlay1986automating,date2018paperprinting,o2015designscape,kang2021metamap,liu2018data,ngoon2021shown,dayama2020grids,zhao2020iconate,kim2022stylette,swearngin2020scout,tanner2019poirot,huang2019swire,zilouchian2015procid,kim2021winder,koch2020semanticcollage,bunian2021vins,kumar2011bricolage,ritchie2011d}\\
    \multirow{16}{*}{} & \multirow{3}{*}{} & Design requirement &\multirow{3}{*}{} &\cite{chen2021cocapture,zheng2019content,wang2020learning,roth1994interactive,guo2021vinci,o2018charrette,kikuchi2021constrained,chen2019gallery,o2015designscape,choi2018fontmatcher,lopes2020adea,lowgren1992knowledge,tan2012development,dayama2020grids,swearngin2020scout,zilouchian2015procid,kumar2011bricolage,ritchie2011d}\\
    \multirow{16}{*}{} & \multirow{3}{*}{} & In line with trends &\multirow{3}{*}{} &\cite{lee2020guicomp}\\
    \cline{2-5}

    \multirow{16}{*}{}& \multirow{4}{*}{Audience} & Favored by the public &\multirow{4}{*}{8.71\%} &\cite{lekschas2021ask,yen2020decipher,kang2018paragon,foong2021crowdfolio,lee2020guicomp}\\
    \multirow{16}{*}{}& \multirow{4}{*}{} & User-friendly &\multirow{4}{*}{} &\cite{kusano2013scenario,lowgren1992knowledge,lekschas2021ask,suleri2019ui,yen2020decipher,kang2018paragon}\\
    \multirow{16}{*}{}& \multirow{4}{*}{} & Reducing errors &\multirow{4}{*}{}&\cite{guo2021vinci,lowgren1992knowledge,lekschas2021ask,yen2020decipher,kang2018paragon}\\
    \multirow{16}{*}{}& \multirow{4}{*}{} & Considering the target audience &\multirow{4}{*}{} &\cite{zheng2019content,maudet2017beyond,o2018charrette,zhao2018characterizes,bylinskii2017learning,suleri2019ui,yen2020decipher,foong2021crowdfolio,lee2020guicomp}\\
    \cline{2-5}
    
    \multirow{16}{*}{}& \multirow{5}{*}{Visual} & Guiding the audience &\multirow{5}{*}{11.29\%} &\cite{bylinskii2017learning,lee2020guicomp}\\
    \multirow{16}{*}{}& \multirow{5}{*}{} & Enhancing readability &\multirow{5}{*}{}&\cite{zheng2019content,guo2021vinci,rebelo2021exploring,ross2020wordblender}\\
    \multirow{16}{*}{}& \multirow{5}{*}{} & Balanced &\multirow{5}{*}{}&\cite{zheng2019content,rebelo2021exploring,kim1993providing,dayama2020grids,lee2020guicomp}\\
    \multirow{16}{*}{}& \multirow{5}{*}{} & Unexpected &\multirow{5}{*}{} &\cite{o2015designscape,kang2021metamap,lopes2020adea,dayama2020grids,swearngin2020scout,mozaffari2022ganspiration}\\
    \multirow{16}{*}{}& \multirow{5}{*}{} & Harmonious &\multirow{5}{*}{} &\cite{wang2020learning,guo2021vinci,rebelo2021exploring,zhao2021selective,yang2019artistic,choi2018fontmatcher,lee2020guicomp}\\
    \cline{2-5}

    \multirow{16}{*}{}& \multirow{1}{*}{Individual design style} & Preference &{\color{red} \multirow{1}{*}{8.06\%}} &\cite{guo2021vinci,kim1993providing,lin2021learning,kikuchi2021constrained,kim2022stylette}\\
    \cline{2-5}
    
    \multirow{16}{*}{}& \multirow{2}{*}{Design contents} & Considering the importance of information &\multirow{2}{*}{14.52\%}&\cite{zheng2019content,guo2021vinci,rebelo2021exploring,newman2000sitemaps}\\
    \multirow{16}{*}{}& \multirow{2}{*}{} & In line with the design theme &\multirow{2}{*}{} &\cite{zheng2019content,wang2020learning,guo2021vinci,ross2020wordblender,ueno2021continuous,zhao2018characterizes,chen2019gallery,kang2021metamap,suleri2019ui,ngoon2021shown,zhao2020iconate,fogarty2008cueflik,tanner2019poirot,koch2020semanticcollage}\\
    \cline{2-5}

    \multirow{16}{*}{}& \multirow{1}{*}{Empty} & Empty &\multirow{1}{*}{N/A}&N/A\\
    \cmidrule[1.5pt]{1-5}
\end{tabular}
}
\end{table*}

\subsubsection{RQ2-1: Which Elements of Tacit Knowledge Receive Adequate or Insufficient Coverage?}

Among the elements, the one least addressed by approaches is related to \textit{design content}, while the most is associated with \textit{work patterns} (\autoref{tab:slr_results}). Specifically, \textit{design content} category shows low occupies in both of tacit knowledge instances (\autoref{tab:instance_results}-Element's Contents) and coverage within existing approaches (\autoref{tab:slr_results}-Element's Contents). In contrast, the \textit{inner design element}, which reveals a significant amount of tacit knowledge instances (\autoref{tab:instance_results}-Element's Inner design elements), demonstrates relatively low coverage (\autoref{tab:slr_results}-Element's Inner design elements). Among the reviewed papers, only an average of 22\% of approaches address \textit{inner design elements} such as font, color, and layout. Notably, instances that can be expressed through user experience exhibit a mere 5\% coverage. Tacit knowledge revealed through \textit{inner design element} is characterized by its close connection to intuition (\autoref{fig:Characteristics}-d), and among the four elements, it represents the most challenging tacit knowledge to communicate (\autoref{fig:Characteristics}-b). In this regard, P3 and P5 mentioned that while knowledge related to inner design elements can be adequately demonstrated through a designer's actions, organizing and reusing such knowledge might be difficult for those who do not possess it firsthand.

\subsubsection{RQ2-2: Which Actions of Tacit Knowledge Receive Adequate or Insufficient Coverage?}
In prior approaches, the well-supported action is linked to \textit{utilization}, while the least-supported action pertains to \textit{cognition} (\autoref{tab:slr_results}-Action's Utilization and Cognition). The \textit{cognition} category includes actions such as possess, feel, predict, and judge. Tacit knowledge that can be performed through these actions accounts for 33\% of the explored instances (\autoref{tab:instance_results}-Action's Cognition), yet it exhibits the lowest coverage within the approaches (\autoref{tab:slr_results}-Action's Cognition). Particularly, the feel action is only supported by one of the surveyed papers. This means that tacit knowledge instances such as \textit{``Feeling good inspiration from unrelated sources''} (P1), related to experiencing something in a sense, are not sufficiently supported yet. P1 further explained that sensing design issues, design inspiration, or possessing design sensibility is difficult to acquire. Communicating such knowledge with others is even more challenging. In terms of characteristics, this action showcases the most challenging aspect in terms of expression (\autoref{fig:Characteristics}-b). In contrast, the \textit{utilization} action is highly supported by prior approaches (\autoref{tab:slr_results}-Action's Utilization). Notably, these approaches usually provide support for finding design references ~\cite{huang2019swire} or adapting external resources ~\cite{lee2020guicomp}.

Tacit knowledge that can be utilized in \textit{manipulation}-related actions exhibits a high coverage in prior approaches (\autoref{tab:slr_results}-Action's Manipulation). Specifically, this action occupies a significant portion of the instances (\autoref{tab:instance_results}-Action's Manipulation). Actions like create, adjust, and combine are encompassed, and many approaches use rule-based or generative methods to manipulate designs directly. Instances applicable to these actions are relatively easy to express and showcase a conventional characteristic (\autoref{fig:Characteristics}-b and d). Designers (P2, P4, and P7-P10) explained that the knowledge to create better designs through manipulation actions is relatively easier to obtain since it can be explained to others relatively easily, and somewhat established methods are available.

\subsubsection{RQ2-3: What Purpose of Tacit Knowledge Receive Adequate or Insufficient Coverage?}

Instances have an amount of \textit{visual}- and \textit{audience}-related purposes (\autoref{tab:instance_results}-Purpose's Visual and Audience), while prior approaches predominantly support \textit{environment}-related purposes (\autoref{tab:slr_results}-Purpose's Environment). Despite a significant focus on goals for design requirements or external constraints, \textit{in line with trends} within the environment had notably low coverage (\autoref{tab:slr_results}-Purpose's Environment). P5 emphasized that aligning with rapidly changing graphic design trends is a crucial but challenging aspect of tacit knowledge in graphic design.

On the other hand, \textit{audience}-related and \textit{individual design style}-related purposes exhibit the lowest coverage within the approaches (\autoref{tab:slr_results}-Purpose's Audience and Individual design style).  P1 and P2 explained that understanding the public's choices or anticipating audience behavior can be acquired through only previous design experiences. Furthermore, they also stated such knowledge is challenging to perceive even with related data, and individuals with an innate sense of this aspect exist. From a characteristic perspective (\autoref{fig:Characteristics}-d), instances linked to \textit{audience}-related purpose often demand substantial personal intuition and are simultaneously possessed by professionals. Instances having a purpose of \textit{individual design styles} are relatively scarce in the graphic design domain and existing approaches (\autoref{tab:instance_results}- and \autoref{tab:slr_results}-Individual design style). Although instances that reflect personal preferences and styles exist, expressing this knowledge itself is exceedingly difficult (\autoref{fig:Characteristics}-b), hence its limited coverage within approaches.

%% file: sections/7_design-guidelines.tex
\section{Design Guidelines for Capturing and Applying Tacit Design Knowledge}
In this section, we propose two sets of design guidelines: 1) to capture tacit knowledge from experienced designers, and 2) to support those that lack the knowledge in applying it in their own design processes. Specifically, we aim to provide guidelines aligned with contemporary graphic design practices, predominantly conducted on web and app-based design tools or systems. The main findings from our work can be summarized as three-fold:
\begin{itemize}
    \item Graphic designers primarily use tacit knowledge by doing manipulation- and cognition-related \textit{actions} on inner design \textit{elements} with the \textit{purposes} of creating visually pleasing artifacts and satisfying the potential audience (\autoref{tab:instance_results})
    \item Graphic design tacit knowledge is acquired through experience and practice (learning-by-doing) and is influenced by personal traits, intuition, and context (\autoref{tab:instance_results}-c and d). However, it can be adequately reasoned about, communicated, and expressed through non-verbal channels like design actions (\autoref{tab:instance_results}-b).
    \item Prior approaches have shown low coverage for the \textit{elements} (contents/inner-design elements), \textit{actions} (cognition/manipulation), and \textit{purposes} (visual/audience) of tacit knowledge (\autoref{tab:slr_results}) because this knowledge is more associated with personal intuition and is more specific to the context.
\end{itemize}

Based on these findings, we aim to answer the following research questions:
\begin{itemize}
    \item RQ3. What are the \textbf{design guidelines} for graphic design tools to share, communicate, and use tacit knowledge?
    \begin{itemize}
        \item RQ3-1. What are the design guidelines for \textit{capturing the tacit knowledge} from the designer who possesses the tacit knowledge?
        \item RQ3-2. What are the design guidelines for supporting designers without \textit{tacit knowledge to apply others' tacit knowledge} in their design process?
    \end{itemize}
\end{itemize}
\label{section_rq3}

\subsection{RQ3-1: Guidelines for Capturing Tacit Knowledge of Graphic Design}
The tacit knowledge of graphic designers is primarily revealed through design elements and actions in the design process (P1-P3, P5, and P7-P10), such as creating, arranging, adjusting, combining, or defining (\autoref{tab:instance_results}-Action). We have confirmed that graphic design knowledge is manifested through inner design elements, outer design sources, and design work patterns (\autoref{tab:instance_results}-Element ). Also, tacit knowledge related to visual purposes is relatively difficult to reason about (\autoref{tab:instance_results}-Purpose and ~\autoref{fig:Characteristics}-b). Related to this, participants (P1, P3, and P8) mentioned that designers often intuitively modify designs based on high-level goals, making it difficult to reason about their intentions at the time. 

However, prior work ~\cite{yen2017listen} and P3 emphasized how reasoning is possible through reflection. Specifically, P3 stated that reflecting on the process along with the final result makes it possible to reason about their decisions after the fact. Furthermore, we found that tacit knowledge can be sufficiently explained through various elements (e.g., fine-grained design editing action, references, prototypes, or design process) beyond only verbal channels (\autoref{fig:Characteristics}-b). Based on these findings, we propose the following system design guidelines for capturing tacit knowledge from professional designers who possess it.

\smallskip
\noindent
\textbf{DG1: Action and Element Tracking}
\begin{itemize}
    \item \textbf{Capture}: Continuously record the designer's low-level actions (e.g., adding/modifying elements, adjusting properties, undo/redo) and corresponding design elements (e.g., shapes, images, text) within and beyond the design tool. This includes actions related to importing/adjusting external resources.
    \item \textbf{Data Integration}: Seamlessly integrate tracking across the design tool and external activities to create a unified action and element sequence.
\end{itemize}

\smallskip
\noindent
\textbf{DG2: Pattern Detection}
\begin{itemize}
    \item \textbf{Uncover Hidden Patterns}: Analyze the collected action data to identify recurring patterns in actions or command sequences, revealing the designer's implicit design process.
    \item \textbf{Focus on Sequences}: Identify and analyze frequent action sequences as they are likely to indicate significant design decisions or strategies.
\end{itemize}

\smallskip
\noindent
\textbf{DG3: Intent Interpretation}
\begin{itemize}
    \item \textbf{Metadata Leveraging}: Utilize relevant metadata associated with the designer's workflow to interpret the intent behind the identified patterns.
    \item \textbf{Contextual Understanding}: Consider the design context (e.g., project goals, target audience, design brief) during interpretations of patterns to gain deeper insights into the designer's thinking process.
\end{itemize}

\smallskip
\noindent
\textbf{DG4: Intent Delivery}
\begin{itemize}
    \item \textbf{Real-time Assistance}: When similar action patterns are detected in a designer's workflow, provide the designer with the interpreted intent along with the associated elements.
    \item \textbf{Adaptive Delivery}: Tailor the delivery of intent information based on the designer's preferences and current context.
\end{itemize}

\smallskip
\noindent
\textbf{DG5: Intent Refinement}
\begin{itemize}
    \item \textbf{Prompt for Annotations}: Prompt the designer to annotate their own interpretations or actual intents on automatically detected action patterns or intents. 
    \item \textbf{Multiple Channels}: Allow diverse channels for annotating the intent, including text, visuals, and elements in ~\autoref{tab:instance_results}, to cater to various aspects of tacit knowledge.
\end{itemize}

\smallskip
\noindent
\textbf{DG6: Tacit Knowledge Storage}
\begin{itemize}
    \item \textbf{Structured Representation}: Store captured tacit knowledge instances as structured data, which includes action patterns, corresponding elements, and the final intent (based on both automatic interpretations and designer annotations).
    \item \textbf{Categorization and Labeling}: Organize and categorize the stored knowledge based on design principles, project types, or other relevant criteria to facilitate retrieval and reuse.
\end{itemize}

\subsection{RQ3-2: Guidelines for Applying Tacit Knowledge in Graphic Design}
Learning-by-doing is the most noticeable characteristic of tacit knowledge in graphic design (\autoref{fig:Characteristics}-c). Much tacit knowledge becomes evident when designers perceive and manipulate inner design elements. Therefore, to acquire this knowledge, it is essential to closely observe design patterns (i.g., editing or creating action) and reasons (i.g., intent or design rationale) for using tacit knowledge within the actual design process. According to P5, providing design feedback is a natural and effective form to support these processes. Prior work ~\cite{kotturi2019designers, krishna2021ready, yen2017listen} also explains that tacit knowledge can be conveyed through design feedback. Based on this, we propose guidelines that specify the format, timing, and content of feedback in the design process to help novice designers acquire and apply tacit knowledge more concretely:

\smallskip
\noindent
\textbf{DG1: Predictive Design Assistance}
\begin{itemize}
    \item \textbf{Predicting the Designer's Next Action}: Employ accumulated tacit knowledge from professional designers to predict the designer's next action.
    \item \textbf{Explain Reasoning}: Provide a rationale for suggested actions, drawing upon relevant design principles and previously captured knowledge.
    \item \textbf{Offer Diverse Suggestions}: Recommend actions beyond simple manipulation, such as suggesting relevant design trends and references.
\end{itemize}

\smallskip
\noindent
\textbf{DG2: Design Knowledge Acquisition}
\begin{itemize}
    \item \textbf{Prompt for Intent Reflection}: Ask novice designers to express their intent or rationale behind specific design actions.
    \item \textbf{Showcase Alternative Perspectives}: Present professional designers' potential actions and intents based on captured knowledge, fostering inspiration, and acquiring professional knowledge.
\end{itemize}

\smallskip
\noindent
\textbf{DG3: Self-Evaluation and Critical Reflection}
\begin{itemize}
    \item \textbf{Prompt Design Evaluation}: Encourage designers to critically assess their own work-in-progress, considering its visual impact and potential response from the audience.
    \item \textbf{Focus on Specific Aspects}: Pose targeted questions to stimulate reflection on key design elements like readability, contrast, and visual hierarchy.
    % \item \textbf{Promote continuous improvement}: Guide designers towards refining their design decisions based on self-evaluation and knowledge application.
\end{itemize}

\smallskip
\noindent
\textbf{DG4: Personalized Design Feedback}
\begin{itemize}
    \item \textbf{Analyze Current Design Elements}: Leverage captured knowledge to offer insightful feedback on the designer's current design choices.
    \item \textbf{Identify Patterns and Trends}: Analyze design work patterns to identify areas for improvement and suggest alternative approaches.
    \item \textbf{Encourage Exploration and Experimentation}: Motivate designers to explore various design possibilities and generate diverse alternatives.
\end{itemize}

%% file: sections/8_discussion.tex
\section{Discussion}
Our work probes the characteristics, instances, and prior approaches of tacit knowledge in graphic design to propose design guidelines for capturing and applying tacit knowledge. In this section, we discuss the implications of our findings, limitations, and future work. 

\subsection{Personal Viewpoints Towards Tacitness: Understanding and Supporting}
In the study with designers, we observed that perspectives on designers' tacit knowledge slightly differed across domains or individuals. For example, some designers said color sense was innate, while others thought it was acquired through practice. This implies that each designer could define the characteristics of tacit knowledge differently and this further explains why tacit knowledge is also called personal knowledge ~\cite{polanyi2012personal}. In this regard, the collection and annotation of tacit knowledge could be conducted at a larger scale and diversity in terms of domain, experience, or work position. This could reveal more general distributions regarding the characteristics of the instances and the diversity of perspectives by domain or level of experience. Since different characteristics would be assigned to the same instances, this could allow for investigating diverse viewpoints toward tacit knowledge. This direction of future work will lead to a more in-depth understanding of tacit knowledge in graphic design, and reveal future directions for the design of more detailed or personalized system support for tacit knowledge.

\subsection{Creativity from the Perspective of Tacit Knowledge}
Instances related to design inspiration, like ``The ability to draw \textit{good motifs or inspiration} from unexpected sources'' (P6), revealed varied perspectives on creativity-related tacit knowledge. Some designers (P1, P8) view it as internalized and innate, while others (P2, P4, and P6) believe it naturally develops through diverse projects. Instances aiming for unexpectedly creative designs (\autoref{tab:tacit-characteristics-list}-Purpose, Visual category) require high intuition and expertise, described as tacit knowledge needing both experience and personal intuition. All designers mentioned it is challenging to explain creativity, but when asked about capturing and reusing creative knowledge, P4 and P6 affirmed its feasibility. They highlighted the importance of tracking observations for creativity sources that lead to inspiration inside and outside the design process. P6 suggested that although explaining real-time creative thinking might be challenging, reflecting on observed elements and personal experiences can facilitate the description. Therefore, despite creativity-related knowledge's high tacitness, we envision a future where designers can share their design creativity as knowledge by capturing and connecting encountered information.

\subsection{Towards Over-the-Shoulder-Learning with Systematic Approaches}
One of the major characteristics of tacit knowledge is that it is a professional’s knowledge ~\cite{koskinen2003tacit, boiral2002tacit, leonard1998role}. Although it depends on the instance, ~\autoref{fig:Characteristics} shows that a high proportion of instances were considered to be characterized as professional knowledge except for the instances of individual design style-related purpose. The most common method for learning design knowledge is through apprenticeship or design studios. While providing these learning methods online could benefit large populations of remote novices, these methods are not suitable for existing forms of online learning, such as video-based learning. This is due to how design knowledge learning and tacit knowledge acquisition necessitate the learner to learn by doing (i.e., designing), instead of only watching a teacher explain the process. In that sense, our work could guide the design of future ``\textit{over-the-shoulder learning}''~\cite{twidale2005over} support systems that follow the learners' design process and provide suggestions that help them learn about and follow tacit knowledge.

\subsection{Generalization to Other Domains: Coding, Writing, and Research}
In this work, we explore and investigate tacit knowledge in graphic design. Our research method can be applied to other domains, such as coding, writing, and research, but tacit knowledge will likely exist in different forms depending on the field. In each domain, future work can identify specific tacit knowledge instances and characteristics first. Further, future researchers can follow our methodology of mapping instances to existing approaches to explore what sort of knowledge is covered by prior work and to reveal under-explored tacit knowledge. For example, in the research domain, there might be tacit knowledge about how to search for references, the order in which to read a paper, and scientific writing skills on how to compose paragraphs during writing. By first revealing what instances of tacit knowledge exist in the research process, future work could follow our research process to then define guidelines for capturing and applying the identified tacit knowledge, and, finally, design systems that employ these guidelines to help early-stage researchers acquire these pieces of tacit knowledge.

\subsection{Limitations and Future Work}
Our work has several limitations. First, although we collected 123 instances of tacit knowledge, it might not be representative enough because the number of designers was ten. A larger number of participants could be recruited and, thus, a larger number of instances could be collected in future work. Second, the recruitment could be expanded for each domain to identify more distinguished features of tacit knowledge in each domain. Third, more various instances of tacit knowledge could be explored and identified in actual design tasks. In our interview study, professional designers replied that the tacit knowledge instance is deeply embedded in their design process. Fourth, the unexplored grounding related to tacit knowledge can be further explored from different viewpoints. The HCI field is highly interested in supporting abstract and tacit concepts like Professional Vision~\cite{beaudouin2023color} and Creativity Support~\cite{li2023beyond} through a tool. In this sense, future work can explore the perspective of what definitions and purposes tacit knowledge is used within the tool. Lastly, if we can capture various instances of tacit knowledge in graphic designers' actual processes, we can adopt further analysis schemes beyond element, action, and purpose. For example, to analyze when tacit knowledge is employed, additional specific contextual dimensions could be introduced, such as the stages of the design process where tacit knowledge is manifested (e.g., design ideation) or the forms of design tasks (e.g., design collaboration). In this regard, future studies could uncover more diverse types of tacit knowledge by collecting data through in-situ design tasks and adopting additional analysis schemes.

%% file: sections/9_conclusion.tex
\section{Conclusion}
This work investigates the tacit knowledge in graphic design by identifying the instances, characteristics, and prior approaches. Through the interview and annotation study with professional graphic designers, this study proposes specific elements, actions, and purposes of tacit knowledge instances in graphic design with their characteristics. Also, systematic literature review and annotation process reveal the less covered elements, actions, and purposes of tacit knowledge by prior approaches that support the graphic design process. Finally, this study proposes design guidelines to capture and support the application of tacit knowledge, considering the characteristics of the instances in graphic design. Our work contributes to the future where tacit knowledge could be actively shared by demystifying tacit knowledge in graphic design.